\documentclass[review]{elsarticle}

\usepackage{booktabs} 
\usepackage{flexisym}
\usepackage[ruled]{algorithm2e} 

\usepackage[caption=false,font=footnotesize,labelfont=sf,textfont=sf]{subfig}
\usepackage{multirow}
\usepackage{array}

\usepackage{algorithmic}

\journal{Journal of \LaTeX\ Templates}

\begin{document}

\begin{frontmatter}

\title{MOF-BC: A Memory Optimized and Flexible BlockChain for Large Scale Networks}


\author{Ali Dorri}
\address{The School of computer science and engineering,  UNSW, Sydney and  DATA61 CSIRO, Australia. }
\ead{ali.dorri@unsw.edu.au}
\author{Salil S. Kanhere}
\address{The School of computer science and engineering,  UNSW, Sydney, Australia. }
\ead{Salil.kanhere@unsw.edu.au}
\author{Raja Jurdak}
\address{ DATA61 CSIRO, Brisbane, Australia. }
\ead{Raja.Jurdak@csiro.au}
\begin{abstract}
BlockChain (BC)  immutability  ensures BC resilience against modification or removal of the  stored data. In large scale networks like the Internet of Things (IoT), however, this feature significantly increases BC storage size and raises  privacy challenges. In this paper, we propose a Memory Optimized and Flexible  BC (MOF-BC) that enables the IoT users and service providers to  remove or summarize their  transactions and  age their  data and to exercise the "right to be forgotten".  To increase privacy, a user may employ multiple keys for different transactions. To allow for the removal of stored transactions, all keys would need to be stored which complicates key management and storage.   MOF-BC introduces the notion of a Generator Verifier (GV) which is a signed hash of a  Generator Verifier Secret (GVS). The GV changes for each transaction to provide privacy yet  is signed by a unique key, thus minimizing the information that needs to be stored.  A flexible transaction fee model and a reward mechanism is proposed to incentivize users to participate in optimizing memory consumption. Qualitative security and privacy analysis demonstrates that MOF-BC  is resilient   against several security attacks. Evaluation results show that   MOF-BC decreases BC memory consumption by up to 25\%  and the user cost by more than two orders of magnitude compared to conventional BC instantiations. 
\end{abstract}


\end{frontmatter}

\section{Introduction}\label{sec:introduction}
BlockChain (BC) is a disruptive technology which has attracted tremendous attention from practitioners and academics in different disciplines (including law, finance, and computer science) due to its salient features which include decentralization,  security and privacy, and immutability  \cite{abramaowicz2016cryptocurrency}. In BC, a transaction forms the basic communication primitive that allows two nodes to exchange information with each other. A digital ledger of transactions is shared and synchronized across all participating nodes. Using a consensus algorithm which involves solving a resource demanding  hard-to-solve and  easy to verify puzzle, a secure trusted network is established among untrusted nodes. BC users may choose to employ changeable Public Keys (PK\textsuperscript{+}) as their identity which prevents them from being tracked, thus increasing their  privacy \cite{dorri2017blockchain2}. Multiple transactions are collated to form a block which is appended to the ledger by following the consensus algorithm. Each block includes the hash of the previous block in the ledger. Any modifications to a block (and thus transactions) can be readily detected as the hash maintained in the subsequent block will not match.   The immutable nature of the BC affords auditability of all stored transactions. \par 
BC was first introduced in Bitcoin \cite{nakamoto2008bitcoin}, the first cryptocurrency system. Since then, it has been widely applied to non-monetary applications, e.g. securing and distributing  sharing economy for renting properties \cite{Slock} and securing vehicle to vehicle communications \cite{rowan2017securing}.  Recently, there has been increased interest in adopting BC to address security and privacy challenges of  large scale networks including  the billions of connected devices that form the Internet of Things   \cite{christidis2016blockchains}.    \par 
\subsection{Motivation}
Although  BC offers a secure, private, and auditable framework for IoT, the immutable nature of the BC raises significant challenges.  First, the large number of IoT devices will undoubtedly generate an equally substantial number of transactions which in turn would significantly increase the memory footprint at the participating nodes that store the BC. To give some context, the current Bitcoin BC which comprises 280 million transactions requires 145GB of storage space \cite{BlockChain}. Arguably, an IoT BC would rapidly outgrow the Bitcoin BC.  The nodes that store transactions in the BC, known as miners, are offered a monetary reward which is paid by the transaction generators in the form of the transaction fee \cite{nakamoto2008bitcoin}. In IoT, the costs associated with a perpetually increasing BC would also be prohibitive for the users. Second, IoT applications are likely to have diverse storage requirements. For example, a smart device may provide its data to a Service Provider (SP) for a fixed period (e.g., a one year subscription to a service) and thus the associated transaction record may only be needed for this duration. As another example, a user may  progressively install multiple IoT devices at a facility and may wish to summarize all transactions associated with these devices into a single consolidated transaction representing the entire facility. Such flexibility is not afforded by current BC instantiations wherein transactions are stored permanently and cannot be altered.  Third, permanently storing transactions of all devices of a user in the public BC  could compromise the user privacy  by: i) linking attack \cite{dorri2017lsb}  whereby the identity of the user is exposed by linking multiple transactions with the same identity, or ii)  monitoring the  frequency with which a user stores transaction even if the transaction content is encrypted \cite{apthorpe2017smart}. Consequently, privacy concerns may motivate  users to exercise their right to be forgotten and not store records of certain IoT devices in the BC.  \par 
\subsection{Contributions}

In this paper, we propose a Memory Optimized and Flexible BC (MOF-BC)  that affords greater flexibility in storage of transactions and data of IoT devices.  The aforementioned examples illustrate instances where the user must have full control over the management of the stored transactions in the BC. One can also envisage use cases where the SP may need to exert this control. For example, the power company may install sensors and smart metering equipment on the client's premises but would still wish to exert control over these devices. MOF-BC introduces User-Initiated  Memory Optimization (UIMO) and SP-Initiated Memory Optimization (SIMO)  which allow either the user or the SP where appropriate to remove or summarize stored transactions and to age (compress)   the data of an IoT device which may be either stored within the BC or off-the-chain in the cloud.  However, the SP or user may not need to be burdened with management of stored transactions for all their devices. MOF-BC thus provides a way to offload these functions to the BC network by introducing the notion of Network-Initiated Memory Optimization (NIMO). This is realized by specifying the Memory Optimization Modes (MOM)  in the transaction when it is created. MOM can be of the following types: (i) do not store: transaction is not stored, (ii) temporary: the network removes a transaction after a specific period of time which conceptually is similar to removing a transaction in UIMO and SIMO, (iii)  permanent: transaction is stored permanently, and (iv) summarizable: the network summarizes multiple transactions to a single summary record in the same way as the user or the SP summarizes  transactions in UIMO or SIMO. The removal of stored transactions  is fundamental to implementation of the aforementioned MOM. However, the immutable nature of conventional BCs does not permit this operation since the hash of a block is computed over the contents of all transactions within the block. The MOF-BC addresses this challenge by computing the hash of the block over the hashes of constituted transactions and not their contents. This allows a transaction to be removed from a block without impacting the hash consistency checks. Moreover, the existence of the hash of a transaction that is no longer present in the BC allows for posthumous auditability. \par 
A previously stored transaction can be removed only by the transaction generator, i.e., the node that knows the corresponding PK\textsuperscript{+}/Private Key (PK\textsuperscript{-}), to prevent another node from either maliciously or erroneously removing its transactions.  As noted earlier, a node (user or SP) may choose to change the PK\textsuperscript{+} used in its transactions to enhance its privacy. The node will thus have to store a large number of keys to be able to remove stored transactions at a later point in time, which complicates key management and storage. To address this challenge,  MOF-BC adds   a \textit{Generator Verifier (GV)} in each transaction which is a signed hash of a \textit{Generator Verifier Secret (GVS)}.  The GVS is a secret that is known only to the entity generating the transaction. It can be a string similar to a password or an image or even biometric information such as a fingerprint scan. The GVS changes for each transaction by applying a simple pattern known only to this entity, e.g. incrementing the GVS by a fixed value, which thus makes the corresponding GV unique. This affords the same level of privacy as with a changeable PK\textsuperscript{+}. Moreover, the GVs in all transactions generated by a node are encrypted using a single PK\textsuperscript{+}. Thus, to verify that a node can remove a stored transaction, it only has to furnish the PK\textsuperscript{+} and the  GVS (and the corresponding GV) for that transaction.   \par 
The removal of a transaction requires each participating node to locate this transaction in the BC, which can in the worst case incur a delay of O(N) where N denotes the number of transactions in the BC. To amortize this overhead, rather than removing transactions on an individual basis, MOF-BC processes transaction removals in batches over a periodic Cleaning Period (CP).   To facilitate the removal process, multiple agents are introduced which reduce the packet and processing overhead associated with multiple memory optimization methods available in MOF-BC. To encourage users to  free up space, MOF-BC grants rewards  to  users that do so and  introduces a  flexible transaction fee based on the duration for which a transaction is stored. The rewards can be either used to pay the storage fee of new transactions or exchanged to  Bitcoin.  The increased flexibility and savings in storage have an associated cost of reduced accountability as the transaction content is either removed or consolidated which leads to loss of some information. Multiple benefits and implications of MOF-BC are studied in this paper.  \par 
To analyze the security of MOF-BC,  we  consider six of the most relevant attacks and outline the defence mechanisms employed to prevent them.   We develop a custom implementation of MOF-BC and show that it   decreases the BC memory footprint by up to 25\% and  the cost  that the users are required  to pay to store their transactions by more than two orders of magnitude compared to conventional BC instantiations.  We also provide comprehensive evaluations on the effects of the CP on the BC size.   \par 
\subsection{Paper Overview}
The rest of the paper is organized as follows. Section \ref{sec:background} discusses related work. Section \ref{sec:RAB}   outlines the details of employed memory optimization methods.  Section \ref{sec:cleaning} discusses the removing process.  Section \ref{sec:evaluation} presents evaluation results.  Section \ref{sec:conclusion} concludes the paper.

\section{Related work} \label{sec:background}
\begin{figure}[h]
	\centerline{\includegraphics[width=8cm,height=5cm,keepaspectratio]{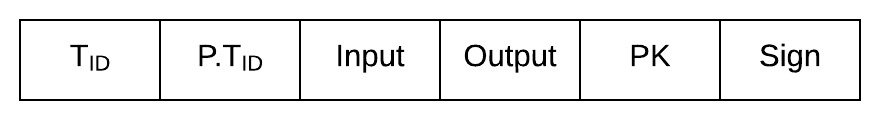}}
	\caption{The structure of a transaction.}
	\label{fig:transaction}
\end{figure}

This section presents a brief overview of BC and outlines related work.  Figure  \ref{fig:transaction} represents the basic structure of a transaction in BC. Note that different instantiations of BC might have some slight variations in the transaction structure. \textit{T\textsubscript{ID}} represents the unique identifier of the transaction which is the hash of all other fields of the transaction. \textit{P.T\textsubscript{ID}} denotes the ID of the previous transaction which effectively establishes the link between successive transactions created by the same node (or entity), thus forming a ledger.  The first transaction in each ledger is known as the genesis transaction.  It is possible that there exist dependences between transactions whereby certain fields generated in one transaction (outputs) are referenced as inputs in another transaction. The inputs and outputs are stored in the \textit{Input} and \textit{Output} fields. In most BC instantiations,  a node changes the public key (PK\textsuperscript{+}) used for encrypting each transaction that it creates as a way to increase anonymity and thus ensures privacy.  The hash of this PK\textsuperscript{+} is stored in the PK field. The reason for storing the hash of the PK\textsuperscript{+} is to reduce the size of the transactions as well as to   future proof transactions against possible  attacks where malicious nodes  may attempt to reconstruct the Private Key (PK\textsuperscript{-}) using the PK\textsuperscript{+} . Finally, the \textit{Sign} field contains the signature of the transaction generator, created using the PK\textsuperscript{-} corresponding to the PK\textsuperscript{+}. \par 

\begin{figure}[h]
	\centerline{\includegraphics[width=5cm,height=5cm,keepaspectratio]{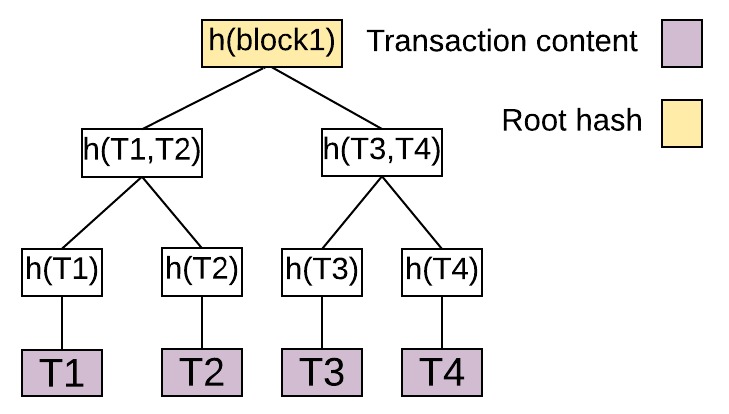}}
	\caption{The structure of the Merkle tree.}
	\label{fig:merkletree}
\end{figure}

All transactions are broadcast to the network. Special nodes, known as miners, verify each transaction by validating the embedded signature using the corresponding PK\textsuperscript{+}.   Next, the existence of the  \textit{P.T\textsubscript{ID}}  is checked. Finally, the other fields in the transaction are verified depending on the specific rules of the BC instantiation. The verified transactions are added to a pool of pending transactions.  Each miner collates pending transactions into a block when the size of the collected transactions reaches to a predefined size known as the \textit{block size}. The miner generates a  Merkle tree  \cite{merkle1987digital}  by recursively hashing the constituted  transactions of the block, which are stored as the leaves of the tree,  as  shown in  Figure \ref{fig:merkletree}.  The root  of the Merkle tree  is stored in the block header to speed up the process of verifying membership of  a transaction in a block.  \par 
The miner  mines, i.e., appends, the block into the BC by following a consensus algorithm, examples of which include Proof of Work (POW) \cite{nakamoto2008bitcoin} and Proof of Stack (POS) \cite{wood2014ethereum}. The consensus algorithm ensures BC consistency between participating nodes as well as randomness among the miners. The randomness   prevents malicious miners from continuously mining blocks, thus increasing BC security. A mined block is broadcast to all nodes. Each node appends the new block to its local copy of the BC after validating the constituent transactions.\par 
In recent years, there has been a growing interest in the adoption of BC technology in IoT. IBM  introduced a new BC instantiation for IoT known as Hyperledger \cite{cachin2016architecture}, which is a permissioned BC, wherein only  authorized nodes can participate in BC.   The authors in \cite{huh2017managing} proposed a new method to manage IoT devices using Ethereum smart contracts. Each device is managed by a contract in Ethereum.  In \cite{sharma2017distblocknet}, the authors proposed a BC-based Software Defined Network (SDN) architecture for IoT. A rule table which defines the rules for access permissions in the network is stored in the BC.  Participating nodes can validate access requests using the BC.  The authors in  \cite{popov2016tangle} proposed a new ledger based  cryptocurrency called IoTA.  By eliminating the notion of blocks and mining, IoTA ensures that the transactions are free and verification is fast. The key innovation behind IoTA is the "tangle", which is essentially a directed acyclic graph (DAG). Before a node can generate a transaction, it has to verify two randomly chosen transactions generated by other nodes. Thus, IoTA  throughput, i.e., the number of transactions validated in the ledger, increases  as the number of participants increases.  \par 
In \cite{shafagh2017towards}, the authors proposed a distributed secure and private IoT data management platform using BC. The data of IoT devices are stored off-the-chain, i.e., in a separate cloud storage, while the hash of the data is stored in the BC. The access permissions for the data is stored in the BC.  The authors in \cite{hashemi2016world} proposed a BC-based multi-tier architecture to share data from IoT devices with organizations and people. The proposed architecture has three main components namely: data management protocol, data store system, and message service. The data management protocol provides a framework for data owner, requester, or data source to communicate with each other. The messaging system is used to increase the network scalability based on a publish/subscribe model.  Finally, the data store  uses a BC for storing data privately. \par

Despite the benefits of the BC in IoT in the above mentioned works, the BC memory footprint still remains an unsolved issue considering the large scale of IoT. As new services are introduced in IoT by SPs and with the passage of time,  the number of transactions generated by  the large number of connected devices that make up an IoT network  will significantly increase.  Consequently, the BC memory footprint will increase infinitely. The authors in \cite{dubovitskaya2017secure, shafagh2017towards} proposed to store the older blocks of the   BC   off-the-chain to address the BC storage challenge. Although this method reduces the resource requirement on the participating nodes, it incurs  delay in querying the transactions from the cloud. Moreover, the issue of ever expanding storage is simply offloaded to the cloud.  Additionally,  as the user data still is permanently accessible by all participants,  the user privacy  is endangered. MOF-BC  tackles these challenges by affording the users flexibility to remove older transactions or consolidate multiple transactions as one and age stored data in the cloud or transactions.  \par 
The authors in \cite{ateniese2017redactable} were the first to propose a modifiable BC that allows users to modify or remove a stored transaction. They use a Chameleon hash function to generate the block  hash, by hashing the  block content, such that a collision can be found in the hash. The modification of BC is performed by one central node or a group of distributed nodes known as modifier(s). The modifiers are aware of a  secret trapdoor key that is used for creating a  collision in the hash of a block such that  $ H(m,\xi) = H(m\textprime , \xi\textprime) $  where \textit{H(x)} indicates the hash of x, \textit{m } is the block content, and $\xi $ is a \textit{check string}.  The  check string is adjusted by the modifier to find a collision in the block hash. \par  
This method faces multiple challenges in an IoT setting.  To modify the content (i.e., transactions) of a chain of blocks, the chain is sent to the modifier(s). The modifier(s)  broadcast the modified chain to all nodes. Each node verifies all  new blocks and replaces the corresponding blocks in its copy of the BC with them. However, this approach is unlikely to scale for the large IoT network eco-system, where one can envision a large number of modification requests, due to significant (processing and packet) overheads associated with the aforementioned steps. Moreover, their approach does not keep the consistency of the transactions before and after modification as the hash of the transaction after modification would not match with the stored hash in the transaction header. Finally, there isn't any provision to remove information from the BC. Conversely,  in MOF-BC only the transaction generator  can modify its transaction. A range of memory optimizations including not storing, removing, and  summarizing transactions and aging of data  are afforded and there is flexibility for initiating these optimizations either by the user or the SP. In addition, the responsibility of these optimizations can be offloaded to the network by generating transactions with specific optimization modes.   MOF-BC introduces multiple agents and a shared indexing service that significantly reduces the associated overheads.  We also propose to use central index database to manage the removal process and introduce multiple rewards to  incentivize users  to free up space in BC.\par 
\section{Optimizing Blockchain Memory Consumption } \label{sec:RAB} 
In this section, we discuss memory optimization methods employed by MOF-BC. We first provide an overview of the MOF-BC framework in Section \ref{sub-sec-overviewofnetwork}. In Section \ref{sub-sec-storagefee}, we introduce the notion of a storage fee as part of the transaction fee to reward the nodes that are involved in  storing the BC for the amount of storage space contributed. Finally, in Section  \ref{sec:mom}, we outline how the underlying mechanisms that form the basis of user-initiated, SP-initiated, and network-initiated memory optimizations   are implemented.

\subsection{An overview of the framework} \label{sub-sec-overviewofnetwork}
\begin{figure}[h]
	\centerline{\includegraphics[width=9cm,height=9cm,keepaspectratio]{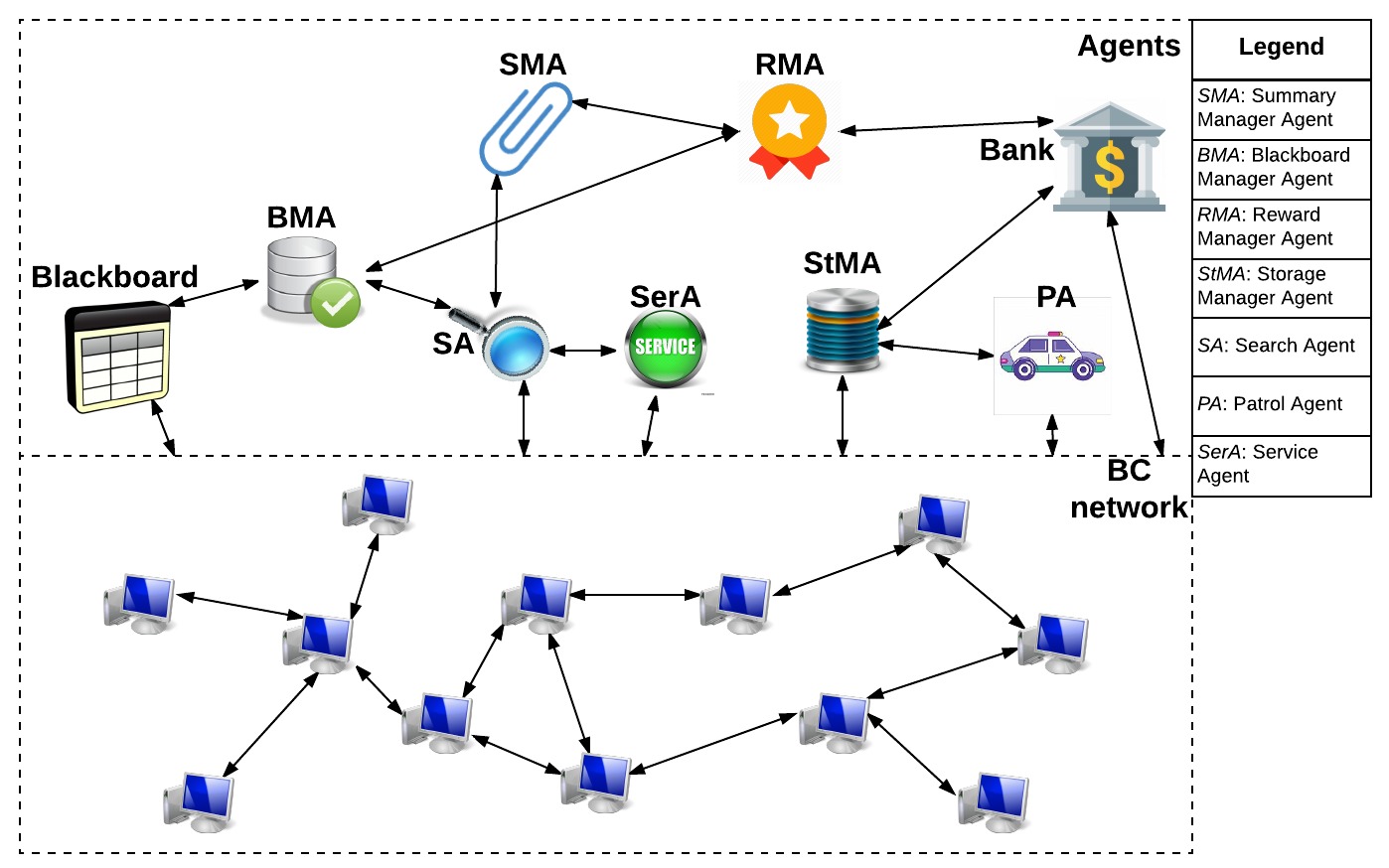}}
	\caption{An overview of the MOF-BC.}
	\label{fig:overal}
\end{figure}

Figure \ref{fig:overal} proposes an overview of the MOF-BC framework. The MOF-BC uses multiple  agents for achieving efficient execution of several key functions. Summary Manager Agent (SMA) manages all the processes related to summarization of multiple transactions into one consolidated transaction.  A Reward Manager Agent (RMA) computes the rewards offered to the nodes that participate in memory optimization and free up memory. The rewards  are sent to a Bank to be claimed by the user. Storage Manger Agent (StMA) collects the storage fees and distributes the collated amount proportionally among the miners. A Patrol Agent (PA) monitors the claims made by the miners by randomly migrating to a miner and checking the storage resources expended in storing the BC.  Blackboard Manager Agent (BMA)  manages a central read-only database  that acts as a repository of information required for bookkeeping e.g., a list of  miners, and a list of transactions paid by rewards. A Service Agent (SerA) is responsible for  processing transaction removals. It maintains an updated version of the BC during the removal process, which is passed on to the miners at the conclusion.  A Search Agent (SA)  searches newly mined blocks for particular, e.g., summarizable transactions and sends them to relevant agents such as the SMA and RMA. The outlined agents may need to generate transactions to action memory optimization, e.g. the SMA generates a summary transaction in place of multiple summarized transactions.   \par 
The agents are partially distributed meaning that multiple replications of each agent are placed in the network.  Synchronization methods such as in \cite{fox1998system} are used to synchronize the replicas.  This prevents the agents from being a bottleneck and reduces  the  (processing and packet) overheads in the network  compared to a fully distributed approach as in \cite{ateniese2017redactable}. \par 

Each agent is identified by a unique PK\textsuperscript{+} which is certified by a Certificate Authority (CA)  to verify the identity of a particular agent. Note that we rely on a centralized approach (i.e., existing public key infrastructure) for this aspect of identity verification. The rest of the functionality is achieved by the distributed BC. \par
\begin{figure}[h]
	\centerline{\includegraphics[width=9cm,height=9cm,keepaspectratio]{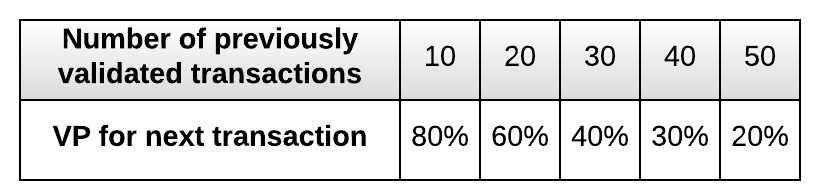}}
	\caption{An example of a trust table in participating nodes.}
	\label{fig:trust-table}
\end{figure}

 
To enhance the network security against malicious agents and reduce the overhead of verifying the transactions  generated by the agents, the miners employ a distributed trust algorithm as proposed in \cite{dorri2017towards}.  The core idea behind the distributed trust algorithm is that the stronger the evidence a node has gathered about  the validity of transactions generated by an agent, fewer of the subsequent transactions received from that agent need to be verified. The miners use a distributed trust table, an example is shown in Figure \ref{fig:trust-table}, which specifies the probability with which a transaction from each agent must be validated (known as  Validation Probability (VP)).   If the transaction is checked and is valid, then  the number of validated transactions for that particular agent is incremented (i.e. this agent is now more trustworthy), thus, the VP for the next transaction from this agent will decrease. If the agent generates an invalid transaction,  the number of validated transactions for that agent reduces. If the malicious agent continues with this behavior, the VP for its transactions will be correspondingly increase.  Note that the transactions always require to be validated with a small probability even if there is strong trust that protects the network against compromised agents.  Due to large number of participants in the BC network, it is with high probability that at least one node will check a new transaction. Thus, the likelihood of detecting invalid transactions is very high. \par 

\subsection{Storage fee}\label{sub-sec-storagefee}
Since the primary goal of MOF-BC is optimizing BC memory storage,  it is important to consider the associated costs and benefits of transaction mining and storage in BC. In conventional BCs, the miner transacts a fee for all transactions within a block as compensation for the resources consumed to mine the block. This fee varies across  BC instantiations.  However, a storage fee is not considered. There is now growing consensus that a fee must be assessed for storage, particularly for large networks like IoT  \cite{Medium}. Billions of devices in IoT will generate a large number of transactions that must  be  permanently stored in the BC and thus requires the miners  to expend significant memory resources for storing the BC.  To incentivize the miners for contributing memory space for the BC storage,   MOF-BC considers  the storage fee as part of the   transaction fee: \par 	\vspace{0.5em}
$ Transaction fee = Mining fee + Storage fee $\par 	\vspace{0.5em}
The mining fee is as was discussed earlier. The storage fee is levied based on the size of the transaction and the duration for which it is stored and is  discussed in Section \ref{sec:mom}. In MOF-BC the minimum size of a transaction is defined as a \textit{page}. For a string \textit{x} we denote its size by $  |x| $. At the minimum, a transaction must include the following fields:  PK\textsuperscript{+}, Signature (\textit{Sign}), hash of the current transaction (\textit{T\textsubscript{ID}}), and hash of the previous transaction (\textit{P.T\textsubscript{ID}}).   Thus, $ |page| = |PK\textsuperscript{+}| + |Signature| + 2|hash| $.  It is assumed that the BC designer sets a pre-defined value for $ |page| $.  The total number of pages required for a transaction  is determined by rounding up to the nearest integer number of pages that can contain the transaction contents.\par 
The  storage fee of all transactions in the network is paid to a Storage Manager Agent (StMA).  The payments to the miners  are made out in periodic time intervals known as \textit{payment periods}. When  miner X joins the BC network, it notifies the StMA of the  amount of dedicated storage space that it is contributing to store the BC, known as \textit{Storage\textsubscript{X}}. Similarly, when miner X   leaves the BC network it notifies the StMA. Thus, the StMA pays the miner only for the duration that it has the BC stored as discussed  below.   The miners may store the comprehensive BC  or a specific part of the BC based on their available resources and requirements \cite{christidis2016blockchains}. At the end of each payment period, the StMA  calculates the cumulative amount of space allocated to the BC in the network (\textit{Store\textsubscript{All}}). The share of each miner of the collected storage fee  (\textit{Share\textsubscript{X}}) is: \par 
$ Share\textsubscript{X} =  Storage\textsubscript{X} * \frac{Fee }{Store\textsubscript{All}} * \frac{Time\textsubscript{X}}{PaymentPeriod}$ \par 
Where \textit{Fee} is the cumulative  storage fees received during the current payment period and \textit{Time\textsubscript{X}}  denotes the duration with the current payment period for which miner X has stored the BC. The StMA may maliciously prevent paying the storage fee to (some of) the miners.  The malicious StMA can be detected by the miners as they would not receive any payment from the StMA. Consequently, the miners isolate the StMA and choose a new one for the network.  All payment claims made by the miners are verified by a Patrol Agent (PA).  The PA is a mobile agent meaning that it can migrate from one miner  to another miner. The PA migrates randomly between miners and examines the memory footprint used at a miner for storing the BC. The PA informs the StMA if a discrepancy is observed in the claims made by a particular miner. No further payment claims are accepted from this malicious miner and  subsequently no rewards are paid out.  \par 
A PA may be compromised. A malicious PA does not report the false claims made by the miners so that the miners will receive payment while they no longer have the BC stored.  This can be detected by the new nodes joining the BC network.  The new nodes  request to download the BC from one of the miners. If the miner rejects their request, they inform the StMA. The StMA then generates a new PA to check the validity of the suspected miner as the old PA is suspected to be compromised. If the new PA detects a malicious behavior  of the miner, then the old PA is removed and the miner is also isolated.  \par 

\subsection{Memory optimization } \label{sec:mom}
In this section, we discuss multiple optimization methods employed by the MOF-BC to optimize the BC memory footprint. As was noted in the examples and arguments in Section \ref{sec:introduction}, either the user or the SP must have absolute control over all stored transactions pertaining to their devices (depending on the use case). To support this,  MOF-BC introduces User-Initiated Memory Optimization (UIMO) and SP-Initiated Memory Optimization (SIMO).  \par 
While the aforementioned optimizations offer immense flexibility, actioning them  requires the user (in UIMO) or SP (in SIMO) to explicitly create  new transactions. For example, to remove a particular stored transaction, the user (or SP) would have to initiated a new remove transaction (details to follow in Section \ref{sec-sub-user-controlled}). These entities may not wish to be encumbered with these actions for all transactions of each and every device under their control. An option to delegate these actions to the network is thus available via Network-Initiated Memory Optimization (NIMO). This is achieved  by specifying the appropriate Memory Optimization Modes (MOM) in the transaction when it is created. \par 
To be able to optimize the BC memory footprint, we introduce the following additional fields to the transaction format: \par 
$ GV  || MOM || MOM-Setup || Pay-by-reward $\par 
GV is used by the UIMO and SIMO as discussed in Sections \ref{sec-sub-user-controlled} and \ref{sec-sub-sub-SP-controlled}. MOM and MOM-Setup fields are used to identify the MOM and its configuration used for the NIMO as disccussed in \ref{sec:network-initiated}. The last field is used to pay the storage fee by accrued rewards as per discussions in \ref{sec-sub-user-controlled}.  \par 
Table \ref{tab:summary-ISMP} summarizes different optimization methods which are outlined in greater detail in the rest of this section. 

\begin{table}[h]
	{
		\caption{Summary of the different optimization options.	\label{tab:summary-ISMP}}
		\centerline{\includegraphics[width=13cm,height=9cm,keepaspectratio]{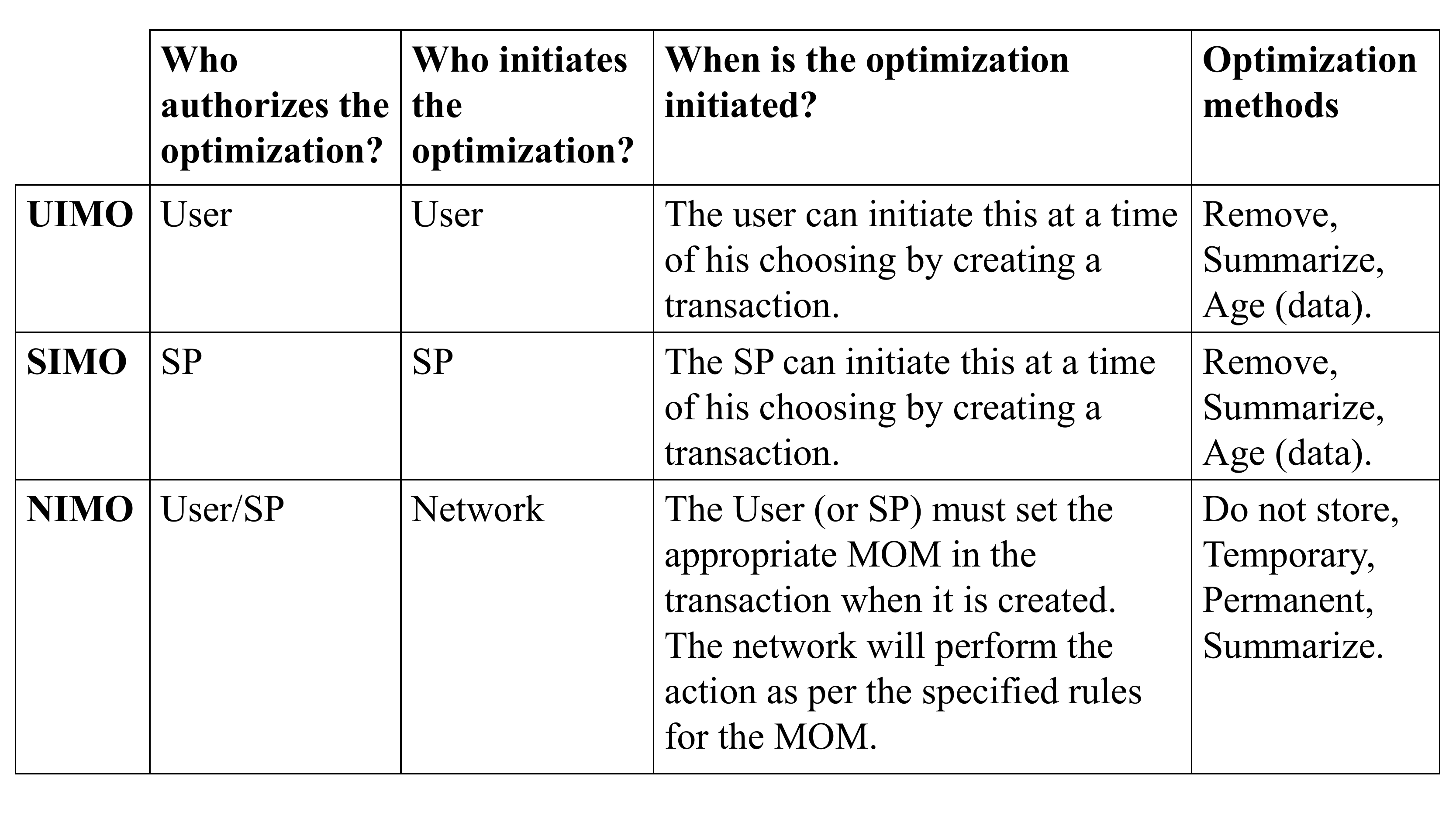}}
		\vspace{-0.7cm}
	} 
\end{table}
\subsubsection{User-Initiated Memory Optimization (UIMO)}\label{sec-sub-user-controlled}
 Before we describe UIMO, we first discuss a new concept introduced for enabling this functionality  known as \textit{ Generator Verifier (GV)}. When a user wishes to optimize a stored transaction, he must first present evidence that the transaction was created by him. The PK\textsuperscript{+} used to generate the transaction is sufficient to prove ownership. However in the IoT context, users may choose to change the PK\textsuperscript{+}  used for different transactions to protect their privacy (as discussed earlier in Section \ref{sec:introduction}). Management and storage of a potentially large number of keys is bound to become an issue.   To address this challenge, we introduce the notion of a  Generator Verifier  (GV). The GV   in all transactions generated by a user is encrypted using a unique PK\textsuperscript{+} (known as GV-PK\textsuperscript{+}), thus eliminating the need for managing multiple keys.  For transaction \textit{i},  GV\textsubscript{i}  is  calculated as the   signed  hash of  the P.T.ID\textsubscript{i} and a  \textit{Generator Verifier Secret  (GVS)}. The GVS   is  a secret known only to the user and can  be any type of data, e.g., a string similar to a password or an image or biometric information such as fingerprint scan. To prevent an attacker from guessing the GVS, the same recommendations for creating strong passwords apply.   The GV value for each transaction is unique even if the same GVS is used in  multiple transactions as the P.T.ID is unique in each transaction.  To further enhance the security,  a user has the option to change the GVS for each transaction by applying a fixed pattern to it. For example, adding a fixed value to it. In this instance, the user would need to remember this unique pattern along with the first used value of the GVS.  Once a transaction with GV is stored in the BC, the user can perform the following memory optimizations at a later time: \par 

\textbf{1) Removal:} The user can remove its stored transactions, to either optimize the BC memory or enhance its privacy, by   generating a \textit{remove transaction}.  To remove a stored transaction, the  user has to  prove that it  has previously generated that transaction. To do so, the user  must include  the hashes used to generate the GV, i.e., the GVS and the P.T.ID, of the  transaction to be removed  and  GV-PK\textsuperscript{+} in the remove transaction. The  ID of the transaction to be removed becomes the input of the remove transaction. To prove that the user knows the PK\textsuperscript{-} corresponding to the GV-PK\textsuperscript{+}, the user signs the remove transaction with this key.  This transaction is subsequently broadcast to the network. \par 
On receipt of the remove transaction (say X),  a miner verifies if the generator of X   is  the  generator of the transaction  that is marked for removal in X (say Y). This verification is conducted using the GV as outlined in Algorithm \ref{algo-GV-verification}.  Since GV is a signed hash, the miner decrypts the GV in Y  using the GV-PK\textsuperscript{+} in X (lines 1,2 Algorithm \ref{algo-GV-verification}). Next, the miner verifies if the hash of  the GVS and P.T.ID in X  matches with the GV in Y (lines 4,5 ). Finally, the miner verifies the signature in X using the   GV-PK\textsuperscript{+} (lines 7,8). This ensures that the generator of X  knows the corresponding PK\textsuperscript{-} to the GV-PK\textsuperscript{+}. The verified transaction is mined into the BC. \par

\begin{algorithm}[h]
	\caption{The process of verifying GV.}
	\begin{algorithmic}[1]
		\algsetup{linenosize=\tiny}
		\scriptsize
		\renewcommand{\algorithmicrequire}{\textbf{Input:} }
		\renewcommand{\algorithmicensure}{\textbf{Output:}  }
		\REQUIRE remove transaction (X),  transaction to be removed (Y)
		\ENSURE  True or False
		\\ \textit{Verification} :
		\IF {(X.GV-PK\textsuperscript{+}\textit{not decrypt} Y.GV }
		\RETURN False; 
		\ELSE
		\IF {H(X.GVS + X.P.T.ID) != Y.GV }
		\RETURN False ; 
		\ELSE 
		\IF {X.GV-PK\textsuperscript{+} \textit{redeem} X.Sign }
			\RETURN True; 
		\ENDIF
		\ENDIF
		\ENDIF
		
	\end{algorithmic}
	\label{algo-GV-verification}
\end{algorithm}

The removal of Y requires each miner to locate this transaction in the BC, which  in the worst case incurs a delay of O(N) where N denotes the number of transactions in the BC. To amortize this overhead, the miners process transactions removal in batches over a periodic Cleaning Period (CP). A detailed discussion about batch removals is presented in Section   \ref{sec:cleaning}. \par 
To encourage users to optimize BC memory consumption, MOF-BC rewards  users that  reduce the BC memory footprint  by  removing   their transactions.  The reward allocation process is performed  by the  \textit{Reward Manager Agent (RMA)}. The RMA would need to search each newly mined block to find a remove transaction. Other agents would need to similarly search for new transactions of a particular kind (e.g. the SMA would need to search for summarization transactions). To amortize these overheads, a dedicated Search Agent (SA) is designated to search newly mined blocks for transactions related to memory optimization and send references for these to the relevant agents.  The agents may randomly check whether the SA sends all relevant transactions to them by searching newly mined blocks and compare the relative transactions with the ones sent by the SA.  For  removed transaction \textit{Y}, the reward value is calculated as: \par 
$Reward = Y.pages - X.pages $\par 
Where \textit{X} is the remove transaction. Each transaction should be only rewarded once, even if it is mined in multiple blocks by multiple miners. To ensure this, the RMA  records the GV of transactions that have  been rewarded.   The RMA  sends the GV and the corresponding amount of reward  to a Bank which collects all information of the rewards. Sending only the reward and the GV to the bank ensures user anonymity.  If a node does not receive its rewards, i.e., the RMA is compromised, then the node informs the rest of the network to change the RMA as per discussions in Section \ref{sec:security-analys}.   \par 
The user, i.e., the rewardee, can use his accrued rewards in two ways: i) Exchange to Bitcoin: A user can ask the bank for his rewards to be converted to Bitcoin at the current exchange rate, ii) Pay storage fee of  new transactions: Recall that the  transaction header contains a pay-by-reward field. If this field is set in a transaction, then it implies that the  corresponding storage fee will be paid by earned rewards. The rewardee should send a corresponding redeem transaction to the bank which contains the ID of the new transaction and  GV-PK\textsuperscript{+} and hashes used to construct the GV corresponding to the accrued reward. The bank verifies the redeem transaction as discussed in Algorithm \ref{algo-GV-verification}. If verified, the bank sends the ID of the new transaction to the BMA  which in turn notifies the miners and  stores the ID of the new transaction in the blackboard.\par 

\begin{figure}[h]
 	\centerline{\includegraphics[width=13cm,height=8cm,keepaspectratio]{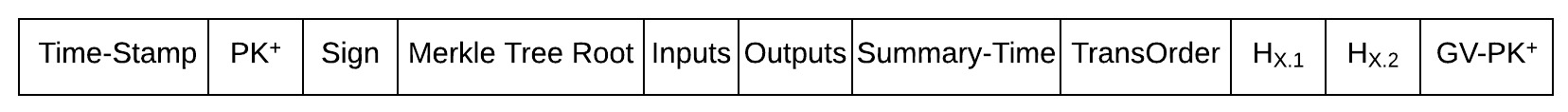}}
 	\caption{The structure of summary transaction.}
 	\label{fig:summary-transaction}
 \end{figure}

\begin{figure}[h]
	\centerline{\includegraphics[width=10cm,height=10cm,keepaspectratio]{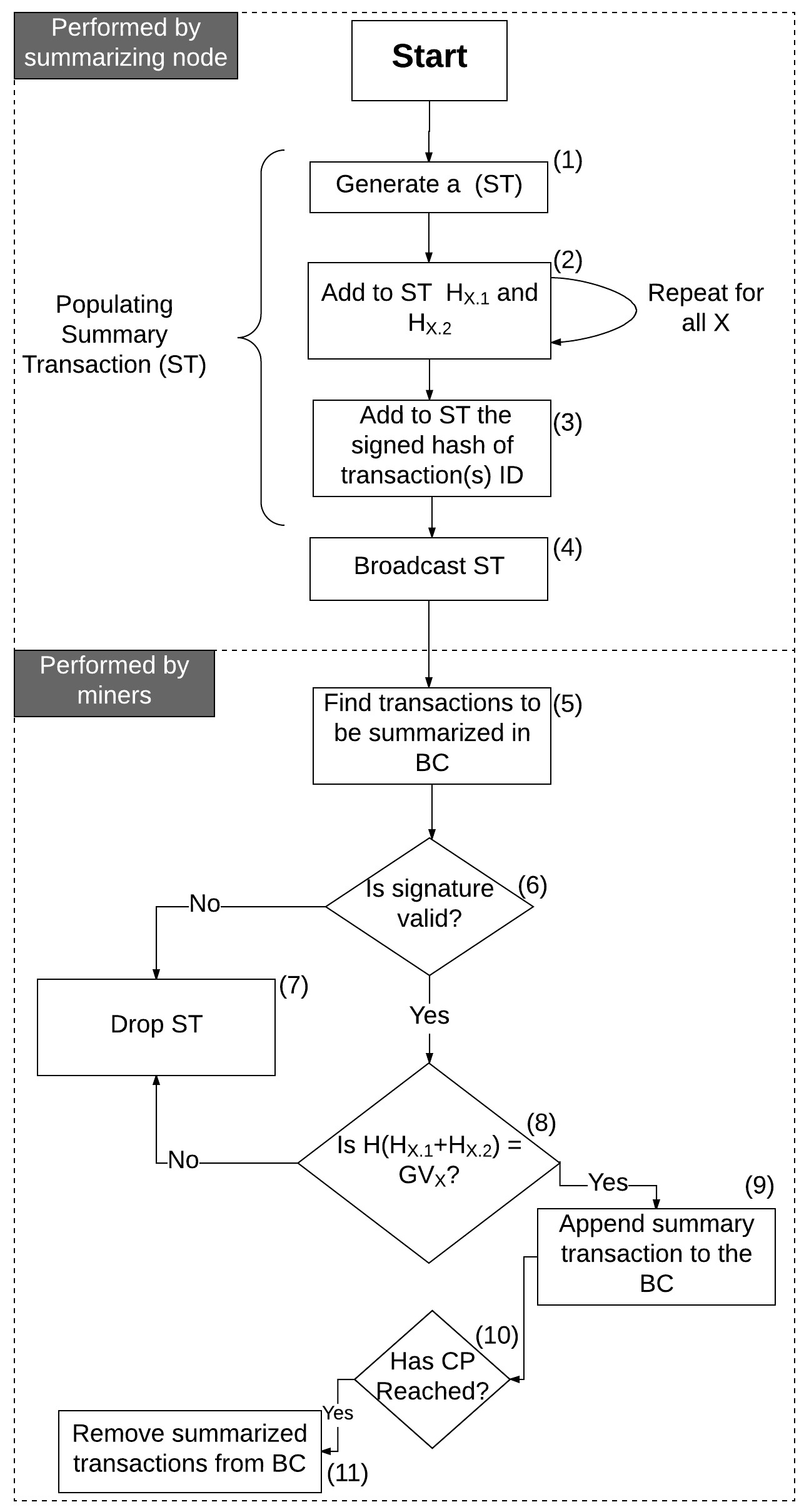}}
	\caption{The process of user-initiated summarization.}
	\label{fig:summary-process-node}
\end{figure}

\textbf{ 2) Summarize: }   As was noted in the examples and arguments in Section \ref{sec:introduction}, IoT users can summarize their transactions into a   single consolidated transaction to optimize BC memory while maintain the auditability of the summarized transactions. The process of summarizing  transactions is shown in Figure \ref{fig:summary-process-node}. To summarize a selected set of transactions, the user creates a \textit{summary transaction}, which is illustrated in Figure \ref{fig:summary-transaction}. \par 
 

\begin{figure}[h]
	\centerline{\includegraphics[width=12cm,height=10cm,keepaspectratio]{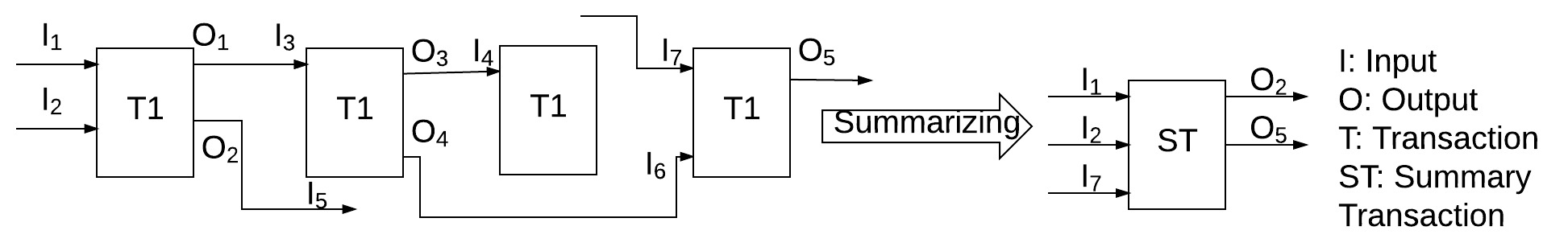}}
	\caption{The input/output of the summary transaction.}
	\label{fig:summary-trans-in-out}
\end{figure}

In Figure \ref{fig:summary-transaction}, the \textit{Time-stamp}  is the time  when the summary transaction is generated. The next two fields are the PK\textsuperscript{+} and signature of the transaction generator. The \textit{Merkle tree root}  is the root of the merkle tree formed by collating all the transactions that are summarized in this consolidated transaction. Recall from Section  \ref{sec:background}, that this data structure makes it is possible to perform posthumous audits, i.e., check whether a transaction belonged to the original group of  transactions that are now summarized.  If the transactions being summarized include multiple inputs and outputs, then including all of them in the summarized transaction would significantly increase its size. As a compromise, we chose to only include the inputs/outputs which  are not used as outputs/inputs of  a transaction in the same summarized group. Consequently, the excluded inputs and outputs can no longer be audited. Figure  \ref{fig:summary-trans-in-out}  illustrates this process. \par 

\begin{figure}[h]
	\centerline{\includegraphics[width=12cm,height=10cm,keepaspectratio]{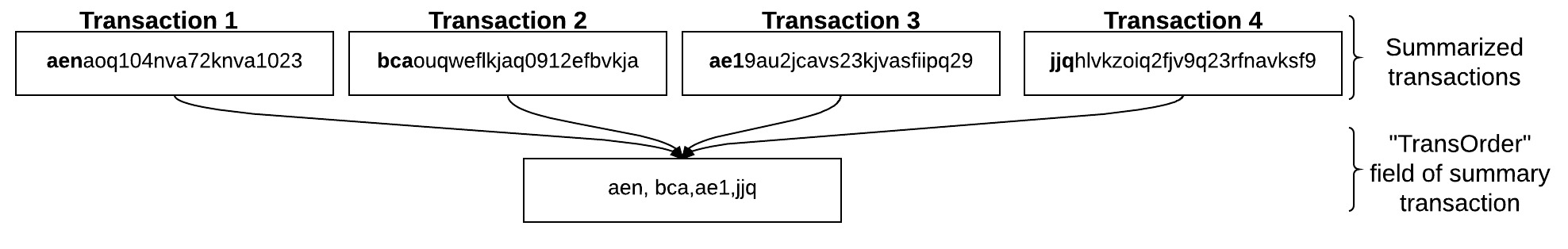}}
	\caption{TransOrder field population.}
	\label{fig:distinct-bytes}
\end{figure}

The time-stamp of each original transaction is stored in the   seventh field, i.e., summary-time.  The  summary-times  are stored in order. However,  the Merkle tree  does not provide any information about the exact order of transactions. To address this challenge,   the smallest number of distinct bytes of the \textit{T.ID} of the summarized transactions (denoted as \textit{d}) are stored in the \textit{TransOrder} field. Thus, by knowing the \textit{T.ID}, the specific order of a transaction  within the pool of transactions  can be found with small overhead. As an example consider the four transactions given in Figure  \ref{fig:distinct-bytes}. The TransOrder field contains  3  bytes of the ID of each transaction as the  first two bytes of  IDs of transactions 1 and 3 are equal. \par 

 The value of \textit{d} is always  $ 1 <= d <= |T.ID-1| $. Given the randomness in the hash outputs it is unlikely that the  hashes for all transactions in the summarization pool  have a large number of similar bytes which justifies using the distinct bytes to reduce the size of the summary transaction compared to storing the original T.ID.   The last three fields in the summary transaction are used to verify the GV and are discussed in the rest of this section. In these fields  X is the index   of the transaction among transactions to be summarized.\par 
 The user initiates the summarization process by populating  a summary transaction (step 1 in Figure \ref{fig:summary-process-node}). To prove that it  has previously generated the transactions that are  to be summarized, the user  stores  the hashes used to generate the GV and the GV-PK\textsuperscript{+} for all transactions to be summarized in the summary transaction  (step 2).  The summarizing node must prove its identity by signing the hash of the ID of transactions to be summarized (step 3). Then, it broadcasts the summary transaction (step 4). \par 
On receiving the transaction, a miner must first  verify if the summary transaction generator has the authority to summarize the listed transactions. This is achieved using  the GV as outlined in Algorithm \ref{algo-GV-verification} (steps 5-9). If the transaction is verified, it is mined in the BC. As was noted in Section \ref{sec-sub-user-controlled}, transaction removals are handled in batches, i.e., at the end of each CP. Thus, the miner will remove all summarized transactions (discussed in detail in Section \ref{sec:cleaning}) from its copy of the BC in the next CP (step 10 and 11).\par  

Recall that MOF-BC rewards users for optimizing BC memory footprint. Similar to how rewards were handled for removing a transaction,   the RMA calculates  the rewards value (\textit{Reward}) for each user  using the following:   \par 	\vspace{0.5em}
$Reward = \sum_{i=1}^{k} t_i .pages - Sum.pages $ \par 	\vspace{0.5em}
Where \textit{k} is the total number of transactions that are  summarized and \textit{Sum} is the summarized transaction.   \par

\textbf{ 3) Aging: } The  data of the IoT devices  is either stored within transactions or in a separate storage (e.g., cloud storage) with the hash of the data  stored in a transaction (i.e., the data is linked to the  BC). To optimize memory consumption, a user may decide to compress the stored data  which is known as aging in the literature \cite{nath2009energy}.  However, applications and services that use the compressed data may  be impacted depending on the extent of the compression particularly when lossy compression is used.  \par 	
In MOF-BC, the users can age their data stored in or linked to a  transaction.  To age data either the user or a third-party, e.g. the cloud storage, passes the original data through an aging function as in  \cite{nath2009energy}.  If a third-party ages data, it must send the aged version of the data to the user for verification purpose.  To prove that the user is the generator of the original transaction, i.e., the transaction that contains  or links to the original version of the data, the GV is included in the aged transaction.  \par 	
The user then broadcasts the aged transaction to the network. The miners verify the transaction by verifying  the GV of the original transaction as outlined in Algorithm \ref{algo-GV-verification}. After verification, the aged transaction is mined in BC.  On receipt of the aged transaction, miners can remove the corresponding original transaction from the BC in the next CP (see Section \ref{sec:data-removing}).    If the removed  transaction is the input of another transaction, then a redirection flag is set in the hash of that transaction in the block, which implies that   the corresponding  transaction  is updated to a new transaction (i.e. redirected). To ensure that users can  find the reference to this new transaction (i.e.,  the redirected address), we use a shared read-only central  database  known as \textit{blackboard}. Multiple replications of the blackboard exist  to reduce the risk of single point of failure and ensure scalability.  The blackboard is managed centrally by a Blackboard Manager Agent (BMA).   The BMA populates the blackboard with the IDs of the aged transactions that have the redirection flag set.   A malicious  BMA may either do not store or alter the data it receives from the nodes. However, this  can be detected by the node (or agent) that originally generates the data to be stored in the BC.  \par

\subsubsection{SP-Initiated Memory Optimization (SIMO)} \label{sec-sub-sub-SP-controlled}
The core functions offered in SIMO are very similar to those in UIMO, i.e. removal and summarization of transactions and aging of data. In SIMO, the SP is aware of the  GV-PK\textsuperscript{+} and GVS of the GVs used by the devices since it exerts control over them and thus the SP can initiate the aforementioned actions. The P.T.ID differs for each transaction, thus, the SP must generate a unique GV for each transaction that its devices are generating as the SP is the only entity that knows the  GV-PK\textsuperscript{+} and its corresponding GV-PK\textsuperscript{-}. This method will not scale for SPs with billions of devices as they require to response to billion of GV generation requests from their devices. To address this challenge, in SIMO the P.T.ID is excluded from the GV generation and  the GV is simply generated  using  the hash of the GVS.  Using the same hash for a number of transactions leads to the same GV value for them. However, as these transactions are the transactions of the SP and  belong to a large number of users, then the privacy of the users is protected.

\subsubsection{Network-Initiated Memory Optimization (NIMO)}\label{sec:network-initiated}
UIMO and SIMO afford significant flexibility to the user and SP for management of their stored transactions. However, there are certain overheads associated with this flexibility. Actioning any of the functions (summarization, removal and aging) requires the responsible entity to create a new transaction. The SP in particular could potentially be responsible for managing a large number of devices and may not want to be burdened with this overhead for all of them.  Since these new transactions need to be mined into the BC, there could be an adverse impact on the BC throughput, i.e., the number of transactions stored in the BC per second. Finally, the removal of a stored transaction  in UIMO \& SIMO achieves zero memory savings as the corresponding  remove transaction needs to be added to the BC.\par 
To address these challenges, MOF-BC offers NIMO, whereby users and SPs can offload these functions to the network. NIMO offers the following memory optimization modes (some of which are very similar  to the functions undertaken in SIMO/UIMO):\par 
\begin{enumerate}
	\item Do not store 
	\item Temporary
	\item Permanent
	\item Summarizable
\end{enumerate}
The MOM field in the transactions (see Section \ref{sec:RAB})  indicates the optimization mode used by the transaction with a different value identifying each MOM.  The MOM field must be set when  the transaction is generated. \par 
\textbf{ 1) Do not store:} It is not necessary to store all transactions in the BC. A transaction may neither use the output nor be the input of another transaction, e.g., a transaction that is generated by a home owner to monitor the security camera of her home. It is not necessary to store such transactions  if the user/SP perceives no benefit in having an auditable record of the same. Moreover, the lack of such a record also increases user/SP privacy.  Transactions flagged as such are not stored in the BC. \par 
\textbf{2) Temporary:}  Certain transactions between IoT nodes might only need to be valid for a specified period of time known as Time To Live (TTL).  For example, a home owner may grant access to the data from a sensor  to the SP for a year's service.  MOF-BC introduces \textit{temporary} transactions for such cases.   A temporary transaction is  removed from BC (as discussed in   Section \ref{sec:data-removing}) after   TTL specified in the MOM-Setup field of the transaction.  \par 
Recall from Section \ref{sub-sec-storagefee} that the MOF-BC applies a storage fee to all transactions.  To encourage   the generation of temporary transactions, MOF-BC introduces flexible storage fee for user/SP that do so: \par 
$ Storage fee = Pages * TTL * Rate $\par 
Rate is the cost of storing a page for a period of time and can be used as a weight to adjust the transaction fee.   The rate could be progressively increased for transactions that are stored in the BC for a longer time. An estimation of the rate can be made based on the  cost of buying storage media. A 1T SSD storage media costs around 650\$. The optimal lifetime of storage media is 5 years \cite{croman2016scaling}.  In our experiments, a page size is 1Kbyte (see Section \ref{sec:evaluation}). Thus, the cost for storing each page for 5 years is 0.00000065\$.   Although storage cost of an individual transaction is small, with billions of IoT devices generating transactions, these costs will add up significantly. The storage fee of a temporary transaction can be paid by  any rewards accrued as per the discussion in Section \ref{sec-sub-user-controlled}.\par 
\textbf{3) Permanent:}
These transactions are stored permanently in the BC. Note that, all current BC instantiations only support permanent storage of transactions. A fixed storage fee is applied, which is defined based on the application. \par
\begin{figure}[h]
	\centerline{\includegraphics[width=12cm,height=10cm,keepaspectratio]{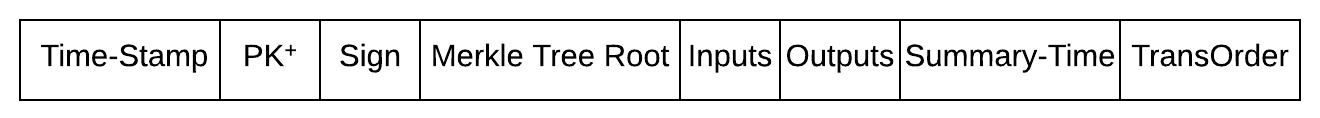}}
	\caption{The structure of NIMO summary transaction.}
	\label{fig:summary-mom}
\end{figure}

 
\textbf{4) Summarizable:} In NIMO, the user/SP must specifically mark transactions that should be summarized when they are created. Since the network handles the summarization process, there is no need for GV as in the case of SIMO/UIMO. Subsequently the structure of the NIMO summary transaction is as shown in Figure \ref{fig:summary-mom}. \par 

Summary transactions are mined as normal into the BC.  At the end of each CP, a  Summary Manager Agent (SMA) summarizes all summarizable transactions in a ledger in a single consolidated summary transaction.  Although the summary transaction is initiated by the SMA, the user/SP first needs to permit its transactions to be summarized by setting the MOM of its transactions as summarizable. This ensures that the user/SP has full control over which transactions should be summarized, while the network specifies when to summarize the transactions to enhance the BC memory optimization. We will further elaborate on the choice of the CP in our experimental evaluations in Section \ref{sec-sub-CPevaluations}. \par
\begin{figure}[h]
	\centerline{\includegraphics[width=6cm,height=6cm,keepaspectratio]{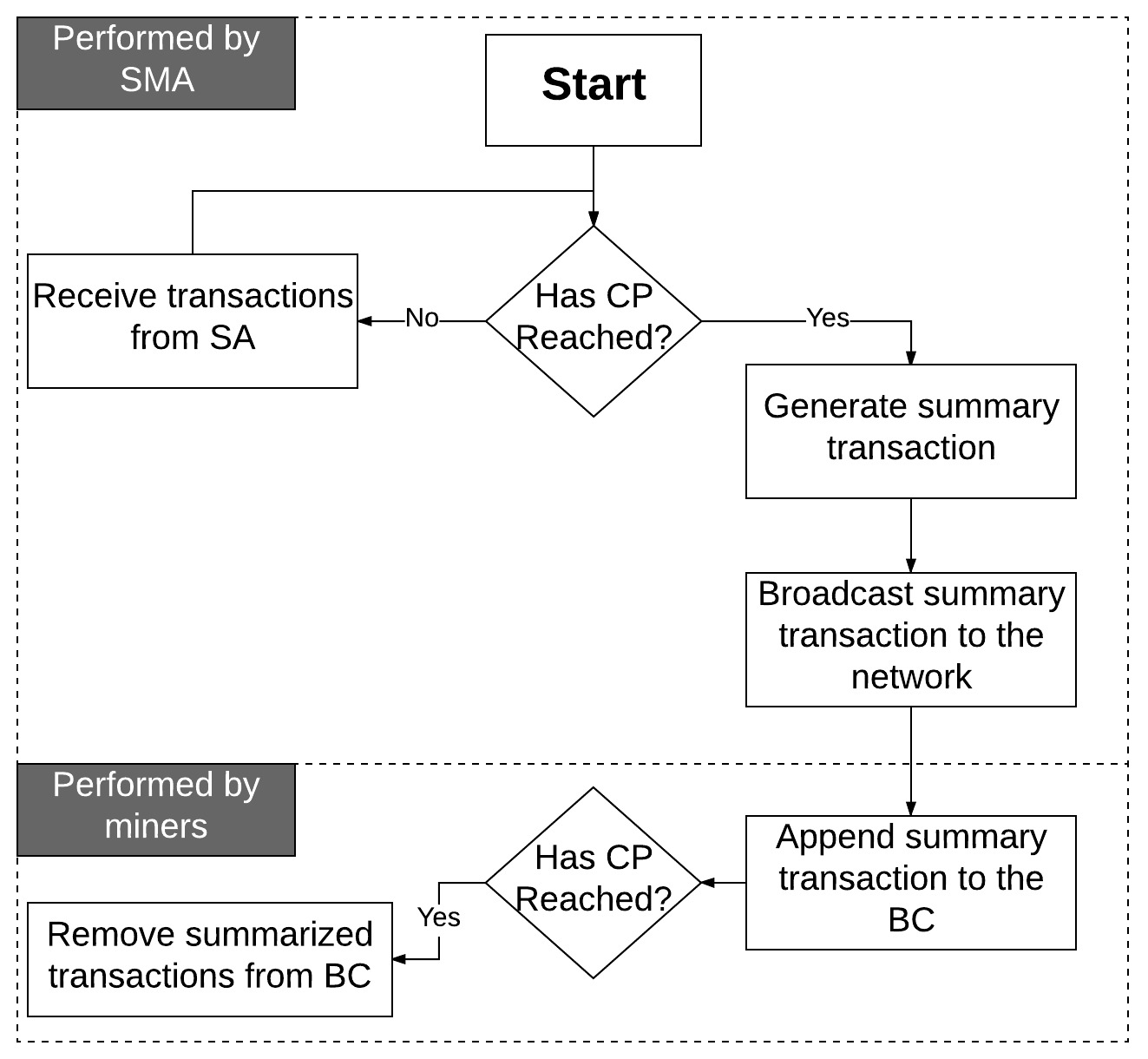}}
	\caption{The process of network-initiated summary transaction.}
	\label{fig:summary-process-MOM}
\end{figure}

 
Figure \ref{fig:summary-process-MOM} outlines the main steps of summarization MOM. The SA scans newly mined blocks for summarizable transactions and sends references to these to the SMA. When the current CP concludes,  the SMA  populates a summary transaction, as outlined in the summarizing process in   Section \ref{sec-sub-user-controlled}, for each ledger that has summarizable transactions.  Next, the SMA  broadcasts the transaction to the network to be mined in the BC. The miners may randomly check the summary transaction by summarizing the summarized transactions and comparing the results with the summary transaction. This protects the network against malicious SMA which generates fake summary transactions. The associated processing overhead with verifying the summary transaction on the miners  is further reduced using  the distributed trust algorithm outlined in Section \ref{sub-sec-overviewofnetwork}.   \par 
As outlined in Section \ref{sec-sub-user-controlled}, the RMA calculates the rewards for each user/SP  that summarized its transactions in the BC.  The total value of rewards (\textit{RewardT})  offered to the owners of summarized transactions   is:  \par 	\vspace{0.5em}
$RewardT = \sum_{i=1}^{k} t_i .pages - Sum.pages $ \par 	\vspace{0.5em}
Where \textit{k} is the total number of transactions to be summarized and \textit{Sum} is the summarized transaction. The share of  node N (\textit{RewardN}) in \textit{RewardT} is:\par 	\vspace{0.5em}
$ RewardN = RewardT* \frac{t_N.pages}{\sum_{i=1}^{k} t_i .pages}$ \par 	\vspace{0.5em}

Although the summary transaction is mined into the BC,  the summarized transactions still need  to be removed. This process is performed at the end of the CP and is discussed in Section \ref{sec:cleaning}.  

\par 

\subsubsection{Summary}\label{sec-summary}
In this section, we summarize key advantages and implications of using   multiple MOMs employed in the MOF-BC in  Table \ref{tab:MOMs}. These arguments are also relevant for SIMO and UIMO, since the underlying functionality for removing and summarizing a transaction in SIMO/UIMO is similar to the temporary and summarizable MOM in NIMO. \par
\begin{table}[h]
	{
		\caption{Discussion about the proposed optimizations. 	\label{tab:MOMs}}
		\centerline{\includegraphics[width=13cm,height=9cm,keepaspectratio]{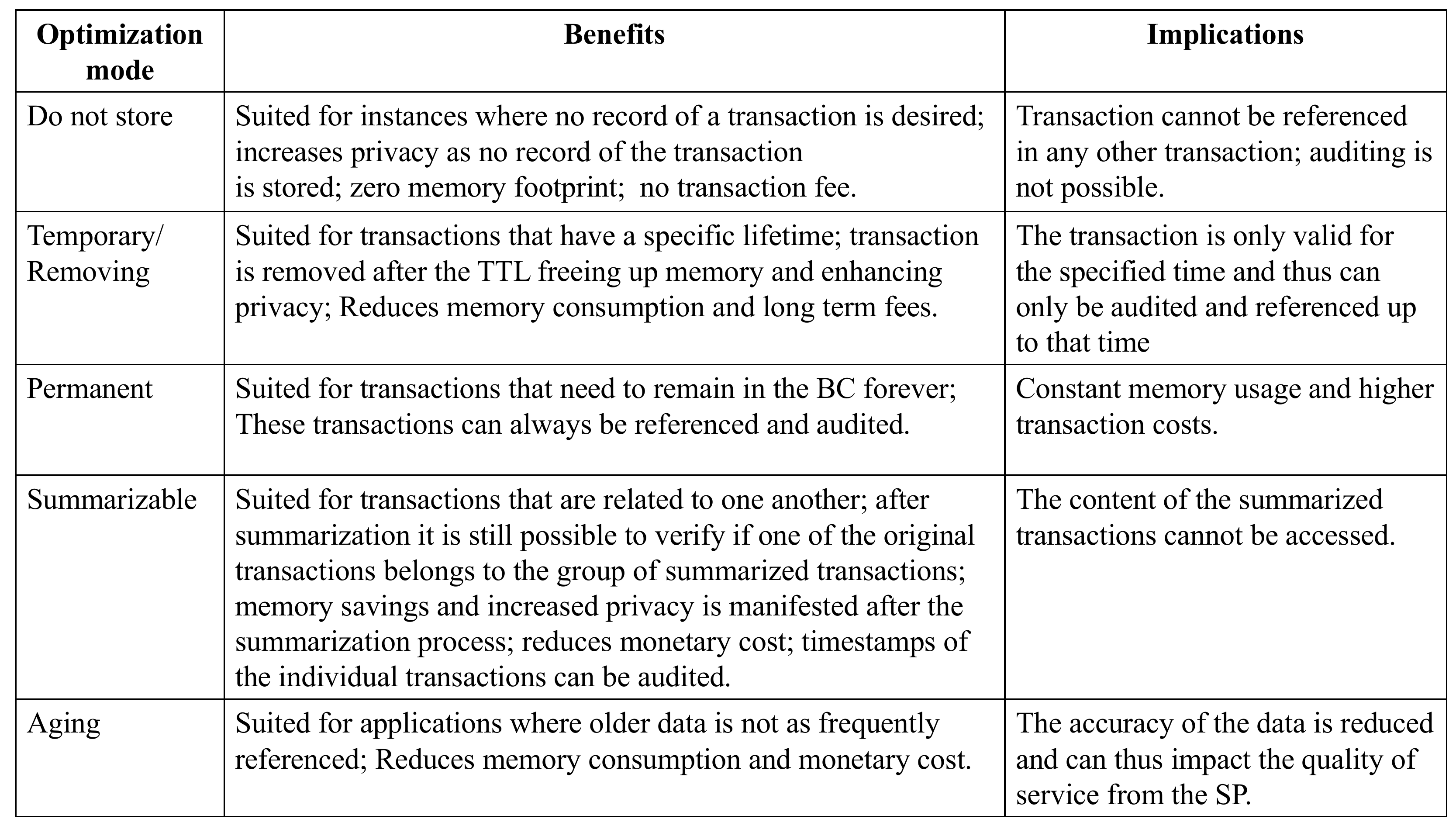}}
		\vspace{-0.7cm}
	} 
\end{table}

\section{Batch Removal of Transactions} \label{sec:data-removing}\label{sec:cleaning}
In this section, we discuss the process for removing  transactions. We also introduce removal of a ledger. Recall that in conventional BCs,  removing a transaction breaks the block hash consistency as the hash is generated over the entire contents of the block as:\par 
$\textit{Block\textsubscript{ID}}= H(T\textsubscript{1}|| T\textsubscript{2}|| ... ||T\textsubscript{k}|| block.header  ) $\par 
Where \textit{k} is the total number of transactions in the block, T\textsubscript{k} is the content of the transaction, and \textit{block.header} is the content of the block header.  To ensure the block hash consistency while  allowing removal of transactions,  in MOF-BC the block hash, i.e., \textit{Block\textsubscript{ID}}, is calculated as:\par  $\textit{Block\textsubscript{ID}}= H(\textit{T.ID\textsubscript{1}}||\textit{T.ID\textsubscript{2}}|| ...|| \textit{T.ID\textsubscript{k}}|| H(block.header))$\par 
\textit{T.ID} is the hash of the transaction content, thus, the transaction content is included in the block hash generation. To remove a transaction, its content is removed from the BC, however,  \textit{T.ID} and \textit{P.T.ID}  remain stored. The   \textit{T.ID} ensures the block consistency while the transaction is removed. \textit{P.T.ID} ensures that the chain of transactions in the ledger is not broken after removing a particular transaction.  \par 
\textbf{Removing a ledger: } In IoT, the users/SPs may demand to remove all information of a particular device.  For example, a device installed in a users home may break. The user may no longer wish to keep a record of the transactions pertaining to this device. Thus, the transaction ledger  is removed using a \textit{ledger-remove } transaction. Normally as outlined in Section \ref{sec-sub-user-controlled}, removing multiple transactions would require the user or SP to provide the GV of each transaction. But since all transactions being removed are chained together in a ledger, it is sufficient to only include the GVS and GV-PK\textsuperscript{+} associated with the genesis transaction of the ledger. This not only reduces the size of the transaction but also simplifies transaction processing. \par 

\textbf{Cleaning Period (CP):} As outlined in removal part of the Section \ref{sec-sub-user-controlled}, the miners process the removal of transactions in batches over a periodic Cleaning Period (CP).   The CP value is application based.   We further elaborate on the choice of the CP and its impact on the BC size and the resources expanded at each miner in the evaluations in Section  \ref{sec:peformance}.\par 
The removal of all transactions at the end of each CP incurs processing overhead on the miners. To reduce this overhead, MOF-BC introduces a  Service Agent (SerA) that handles the removal process and makes its updated version of the BC available for all miners to download, which in turn reduces the processing overhead on the miners. The miners (or some of them) may decide to perform the removals by their own and compare the result with the updated version of the BC available from the SerA to ensure that the SerA is not compromised.  Similar to SMA, distributed trust algorithm is used to decrease the processing overhead. 

\section{Evaluation and Discussion} \label{sec:evaluation}
In this section we provide qualitative security  analyses as well as quantitative performance evaluations. 
\subsection{Security analysis}\label{sec:security-analys}
We first discuss MOF-BC  security and  fault tolerance. It is assumed that the adversary can be any node in the BC network, e.g., miner, SP, agent, or cloud storage. Adversaries are able to sniff communications, create fake transactions, attempt to change or remove stored transactions in BC, and link a user's transaction to each other to uncover the real identity of the user.  However, they are not able to compromise the standard encryption algorithms that are employed.\par 
\textbf{Security: } We consider the following attacks: \par  
\textit{Transaction Removal Attack: } In this attack, the malicious node attempts to remove the transactions  generated by other nodes. Recall that in MOF-BC 3 entities can initiate  the memory optimization: the  user in  UIMO (Section \ref{sec-sub-user-controlled}), the SP in SIMO \ref{sec-sub-sub-SP-controlled}, and the network in NIMO \ref{sec:network-initiated}. To remove the transactions of a user or SP, the malicious node requires: i) The  GV-PK\textsuperscript{+}  to decrypt the signature of the GV which can be gained from the remove requests generated by the true user/SP, ii)  The corresponding PK\textsuperscript{-} to GV-PK\textsuperscript{+}   to sign the  remove request which is only known to the user/SP.  In the worst case, if the attacker somehow finds the PK\textsuperscript{-},  it still requires the GVS to decipher the GV. A brute force attack is unlikely to succeed given that a collision resistant hash such as SHA-3 is employed.   \par 

In NIMO, the specific MOM is indicated within the transaction when it is created by the user or SP making the  transaction   resistant against this attack. \par 

\textit{False Storage Claim:} In this attack, the  malicious miner claims to have  BC stored  to receive incentive from the StMA (see Section \ref{sub-sec-storagefee}). Recall that the PA migrates randomly between the miners that have made claims  and validates their claim by examining the space that they have dedicated to the BC.   To verify the storage claims  of all miners, the PA must visit all miners at least once during the CP. The average frequency of visiting the miners by the PA can be defined by network designers considering the outlined implications.  \par 
\textit{Agent Isolation: } An agent may become completely isolated if all the nodes in its one-hop neighborhood collude with each other. These colluding nodes may  isolate the agent by dropping all transactions or blocks to or from the agent. The aim of this attack is to  prevent normal well-behaving nodes from receiving services offered by the agent. To mitigate the effect of  this attack, in MOF-BC multiple replicas of each agent  are positioned in different places in the network. Thus, in case one of them is isolated, then  other agents can  continue to  provide service.  We elaborate more on the impact of the number of isolated agents on the service provided to the nodes while we discuss fault tolerance at the end of  Section \ref{sec:security-analys}. \par 

\textit{Malicious SP: } A SP may maliciously remove a transaction to disown responsibility. Consider the following example. Alice (the renter of a home) has sent a transaction to Bob (the home owner and thus the SP) alerting him that the fire alarm is broken and requesting servicing. Bob ignores the transaction. A fire breaks out in the home causing significant damage. Bob wishing to shirk responsibility removes Alice's transaction and falsely alleges that she was responsible for the fire as the faulty fire alarm was not reported.   Even if Bob removed Alice's transaction, its hash is still present in the BC. Assuming that Alice has stored a copy of the transaction locally, she can readily verify this and thus implicate Bob. \par 
\textit{Reward sniffing:} The attacker sniffs the communications between the users/SPs and the bank to discover the PK\textsuperscript{+} that  rewards are paid to. The attacker can subsequently  track the user/SP payments and compromise his privacy. The users/SPs encrypt the  redeem request using the PK\textsuperscript{+} of the bank, which ensures only the bank can read the GVs and the PK\textsuperscript{+} to be paid to. Additionally,  the users/SPs can exchange their earned  coin or reward with any other user/SP. Thus, even knowing the PK\textsuperscript{+} of the rewards corresponding to each user/SP does not compromise   user/SP privacy. \par 
\textit{Malicious agents: } The agents  may perform malicious activities in the network. It is assumed that agents are selected to run on nodes which have higher security, e.g. the machines with high resources to perform  the security tasks. However, we conservatively assume that the agents can still be compromised. Table \ref{tab:agent-security}  summarizes the key methods employed by the miners to detect misbehavior of the agents.  Once a malicious agent is identified, the miners isolate it and choose a new agent as a replacement.\par 

\begin{table}
	\caption{The  employed methods to detect misbehavior of agents.}{
		\begin{tabular}  {| p {1cm} | p { 10.5 cm} |}
			\hline
			\textbf{Agent  }     & \textbf{The employed method} \\\hline
			SA  &  The agents that receive transactions from the SA randomly search the new blocks for the relevant transactions and compare them with the  transactions  sent by the SA (see Section \ref{sec-sub-user-controlled}). \\\hline
			SMA & The miners randomly summarize the summarized transactions using the same method as the SMA and validate the received summary transaction (see Section \ref{sec:network-initiated}). \\\hline
			SerA & The miners verify the received blocks by ensuring that the SerA only removed the expired transactions from the BC (see Section \ref{sec:cleaning}).   \\\hline
			RMA & A compromised RMA  can be  detected by the nodes in BC network  as they would no longer receive reward or receive partial rewards. Such nodes inform the rest of the network of the malicious behavior of the RMA (see Section \ref{sec-sub-user-controlled}). \\\hline
			BMA & Similar to RMA, if the BMA is compromised the participating nodes in the BC or the agents that populate the blackboard can detect the misbehavior as the information they've sent to the BMA is not stored in the blackboard (or is changed) (see Section \ref{sec-sub-user-controlled}). \\\hline
			StMA & A compromised StMA can be detected by  the miners as they are not paid for their service (see Section \ref{sub-sec-storagefee}). \\\hline
			PA & The malicious behavior of the PA is detected by the new nodes joining the BC. These nodes request to download the BC from one of the storing nodes. If the request is rejected by the storing node, then the new node informs the StMA. The StMA verifies the PA by generating a new PA controlling the suspected storing node (see Section \ref{sub-sec-storagefee}). \\\hline
	\end{tabular}}\label{tab:agent-security}
\end{table}

\textbf{Fault tolerance: } Fault tolerance is a measure of how resilient an architecture is to node failures. In MOF-BC only the agents are vulnerable to failure as the rest of the  network that forms the BC works distributedly. To enhance the fault tolerance of the  agents, multiple replicas of the agents work collaboratively. Thus, failure of one  will not impact the network. Failure of multiple replicas of an agent may affect the fault tolerance of the MOF-BC. If the total requests generated by the participating nodes in the BC exceeds the cumulative response rate, i.e., the number of transactions being responded, of  replicas, then the participating nodes experience delay in receiving service. 

\subsection{Performance evaluation}\label{sec:peformance}
In this section we present extensive performance evaluations of MOF-BC. We implemented MOF-BC using C++ integrated with crypto++ library and SQLite database on a  MacBook laptop (8 GB RAM, Intel core M-5y51 CPU, 1.10 GHz*4). We assume a network comprised of 900 nodes, each of which has its unique PK\textsuperscript{+}/PK\textsuperscript{-} for generating transactions. Each node generates one transaction per week. Nodes generate different types of transactions and the  network initiates the  memory optimization.   In BC, throughput is the defined as the total number of transactions that can be stored in BC per second. The BC throughput in our implementation is 90 blocks per week, i.e., one transaction for each node per week.  Transactions are organized in blocks such that there are 10 transactions within one block. Since the process of mining blocks is orthogonal to the functionality offered by MOF-BC, i.e., optimizing BC memory footprint, it has not been implemented. Each transaction contains PK\textsuperscript{+}, signature, T\textsubscript{ID}, P.T\textsubscript{ID}, MOM, and MOM-setup (see Section \ref{sec:mom}) fields. \par 

Recall from Section \ref{sec:network-initiated} that in NIMO a large portion of the optimization tasks are performed by the network, thus, increases the (processing) overhead on the network compared to UIMO and SIMO   where the user or SP performs most of the processing for the memory optimization. Additionally, MOMs defined in NIMO overlap with the functionality supported in UIMO and SIMO.  Thus, we only evaluate NIMO since we expect that the associated  overheads    cover the overheads incurred by other two optimization methods. Note that the optimized memory using the aging largely depends on type of data and aging function which are not in the scope of this paper.  \par 

We first evaluate  the benefits (memory footprint optimization and monetary cost) and implications (processing time of handling memory optimization tasks) of using MOF-BC compared to scenario when all transactions are permanently stored. We show that the benefits of MOF-BC far outweight the small overheads. Next, we provide comprehensive  evaluations on memory saved or expended of choosing the CP values. 

\subsubsection{BC size and transaction fee}
\begin{figure} 	
	\begin{center}
		\subfloat[]{\label{fig:evaluations-fee-memory} \includegraphics[width=5.5cm,height=4cm,keepaspectratio]{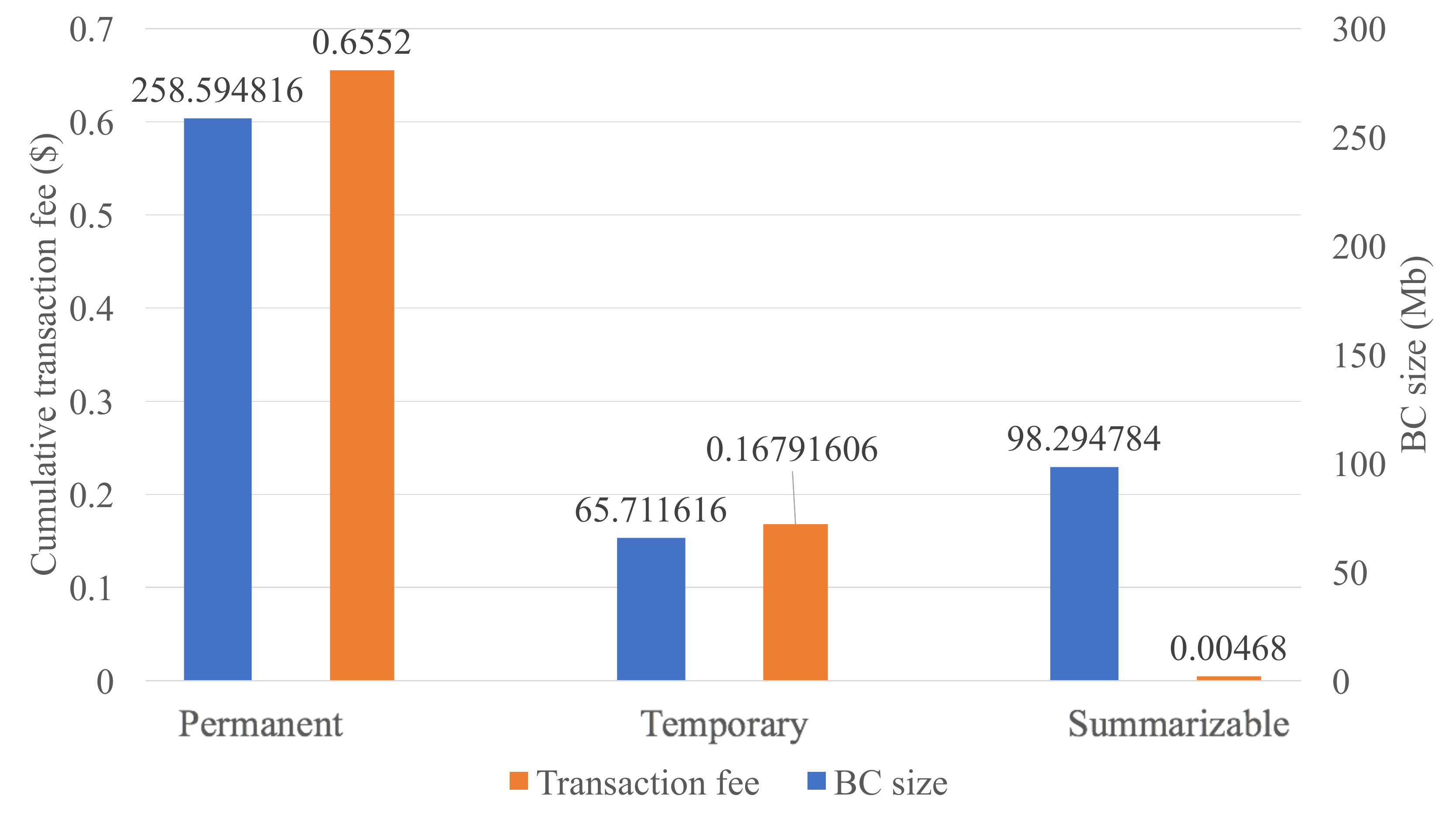} }\hfill
		\subfloat[]{\label{fig:evaluations-processing} \includegraphics[width=5.5cm,height=4cm,keepaspectratio]{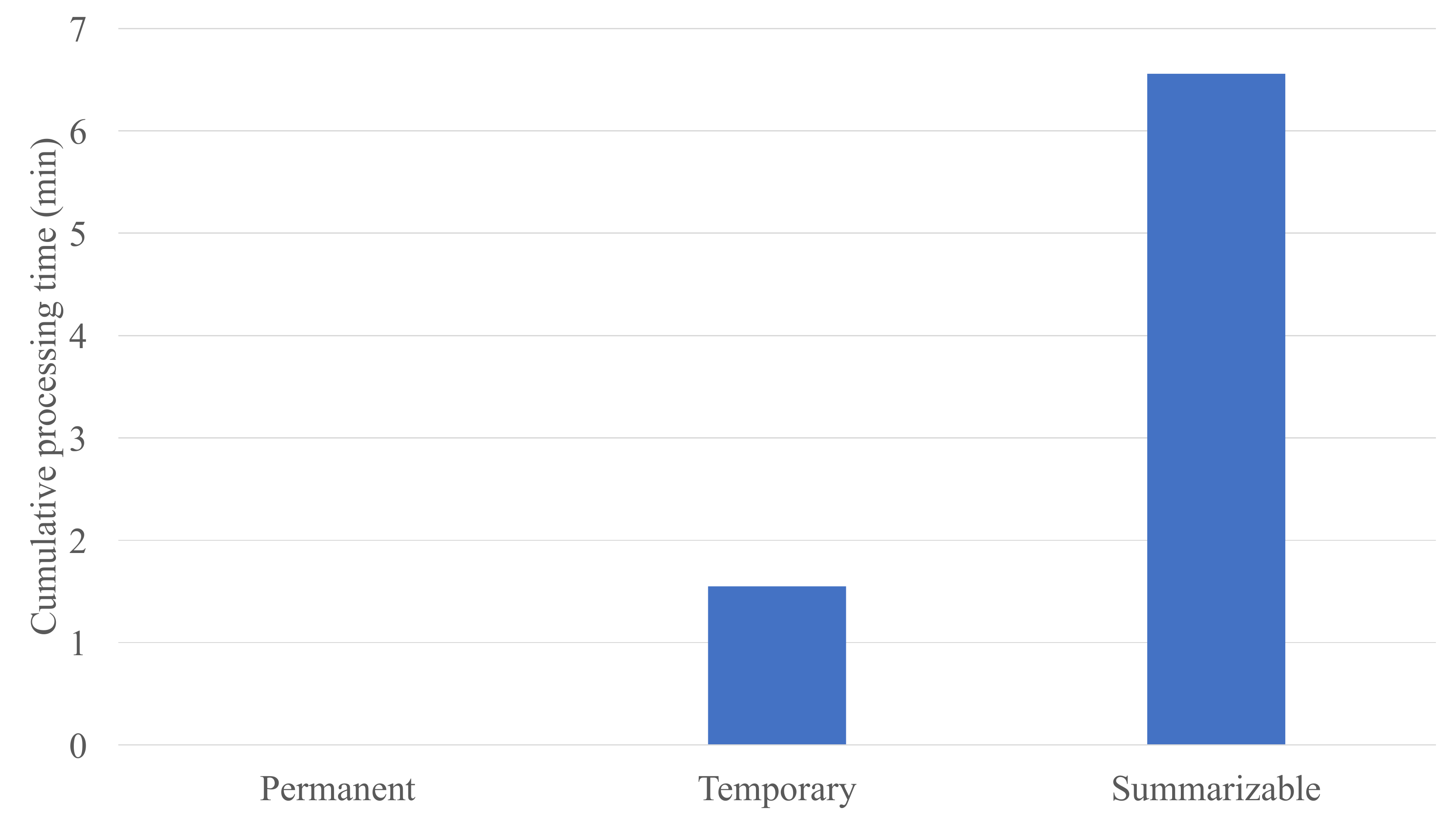} }\\
		\caption{  Evaluation of the a)  cost and memory b)  processing time.\label{fig:evaluation-BCsizevsFee}}
		\vspace{-0.6cm}
	\end{center}
\end{figure}

In this section, we evaluate the impact of multiple MOM on the BC size and the transaction fee that needs to be paid by the users. It is assumed that each node generates one  transaction per week with a cumulative total of 280 transactions. All nodes generate transactions sequentially which are  organized in blocks of 10 transactions resulting in a BC that contains  25200 blocks. The CP is set to 180 weeks. The evaluation metrics are computed once the BC is populated with 25200 blocks.  We use 3 different instances of the BC each using a different MOM, to evaluate each MOM separately. The CP equals to 2 weeks, i.e, 180 blocks. For temporary MOM we form three groups of nodes, each group with 300 participating nodes. All participants of a group use the same value as the TTL which is 26, 52, and 104 weeks for groups 1-3 respectively. Groups sequentially generate one transaction. To measure the transaction fee, we used the estimated storage fee in Section \ref{sec:network-initiated}, i.e. 0.00000065\$ per 5 years.   The cost for storing a temporary transaction for 20 years is considered as the permanent transaction fee which is 0.0000026\$.  Mining fee is not considered in our study as we exclusively wish to consider the storage fee as an evaluation metric.   The implementation results are shown in Figure \ref{fig:evaluation-BCsizevsFee}.  In Figure  \ref{fig:evaluations-fee-memory} the right vertical axis shows the BC size  while the left axis shows the cumulative  transaction fee which is the transaction fee paid by all nodes in the network. Figure \ref{fig:evaluations-processing} illustrates the cumulative processing time incurred for executing the actions associated with the 3 MOM. \par 

Observe that, the permanent MOM  is essentially similar to conventional BCs expectedly has the highest memory footprint. The BC size of temporary MOM is the lowest  as transactions are removed after their TTL expires. With the summarizable MOM, a sizeable fraction of the BC size is attributed to summarized transactions.  Consequently, the BC size is greater than with temporary. However, the summarizable  transaction incurs lower cost compared to the temporary. For temporary transactions,  the node pays a flexible transaction fee for all its transactions.  For summarizable transactions, each node receives a reward at the end of each CP which is used to cover part of the transaction fees, thus reducing the expenses incurred by the user.  In our experiment at the end of the first CP each node earns 1.6  rewards based on the reward calculation formula given in Section \ref{sec:network-initiated}. Following this, in each CP one summary transaction for each node is stored in the BC, thus each node can only store one transaction in the BC within the CP. The storage fee of this transaction can be paid by the earned rewards from the optimizations employed in  the previous CP. Thus, each node only needs to pay for its first two transactions stored during the first CP and the rest are paid by earned rewards.  As shown in Figure \ref{fig:evaluations-processing}, the processing time for summarizable transactions is greater than the temporary transactions. This is because the transactions that are summarized need to be removed and a new summary transaction must be mined into the BC. Note that the measured processing time is the read/write time for updating the database and thus is the HDD read/write time.  
 \par 
We next measure the cumulative monetary cost saved by the network participants using the MOF-BC as well as the monetary cost of running MOF-BC tasks. Thus, we require to convert the processing time incurred by the MOF-BC to  energy consumption and in turn  to monetary cost. To convert the processing time to  energy consumption, we base our calculations on the energy consumptions measured by the authors in  \cite{jaiantilal2010modeling}.  The consumed energy during the load time for HDD is 8.4W. Thus, the cumulative consumed energy for temporary and summarizable MOM is 781 and 3305W respectively. We next measure the cost for energy. Based on the market price, the energy price  is 28.52 c/Kwh \cite{Origin}.  The saved and incurred costs are presented in Table \ref{tab:cost}. The saved cost is the cost that each user saves in transaction fee and is measured by subtracting the  cost that the user pays for using the temporary and summarizable MOM from the cost that the user would have paid for the permanent MOM,  which is essentially similar to the conventional BCs, and the cost users pay using each MOM. The incurred cost includes the cost incurred at the miner for executing the removal process (see Section \ref{sec:cleaning}). \par 
It is evident that the cost saved by the MOF-BC is by far higher than the cost incurred for processing transactions. As the memory footprint of MOF-BC reduces significantly compared to conventional BC instantiations, the miners are required to expend less memory to store the BC and thus can save monetary cost of purchasing and maintaining extra storage. The exact amount of saving depends on multiple factors including  the storage space in the miner, the BC size, and the wasted memory which are application-specific, thus, we are unable to provide any estimation on this additional advantages of the MOF-BC. \par 

\begin{table*}[h]
	\caption{The saved and incurred cost by MOF-BC (\$).}{
		\begin{tabular}  {| p {4cm} |  p {3.2 cm}| p {3.2 cm} |}
			\hline
			& \textbf{Temporary} & \textbf{Summarizable } \\\hline
			\textbf{Saved cost }  & 0.48728394 & 0.65052 \\\hline
			\textbf{Incurred cost }  & 0.000374948 & 0.001586572 \\\hline
			\textbf{Benefit/Cost ratio} & 1300 &  410 \\\hline
	\end{tabular}}\label{tab:cost}
\end{table*}

\subsubsection{Impact of varying the CP} \label{sec-sub-CPevaluations}
Recall from Section \ref{sec:cleaning} that in MOF-BC at the end of each CP each node that stores a copy of BC removes all removable transactions from its BC copy. Additionally,  the SMA generates a summary transaction for all summarizable transactions in a ledger and broadcasts it to be mined into the BC.  We disregard the delay of creating a summary transaction and the time taken for the broadcast to propagate through the network as  these have no effect on the measured metrics. Thus, the summary transactions are immediately mined into the BC after generation.  \par
\begin{figure}[h]
	\centerline{\includegraphics[width=9cm,height=6cm,keepaspectratio]{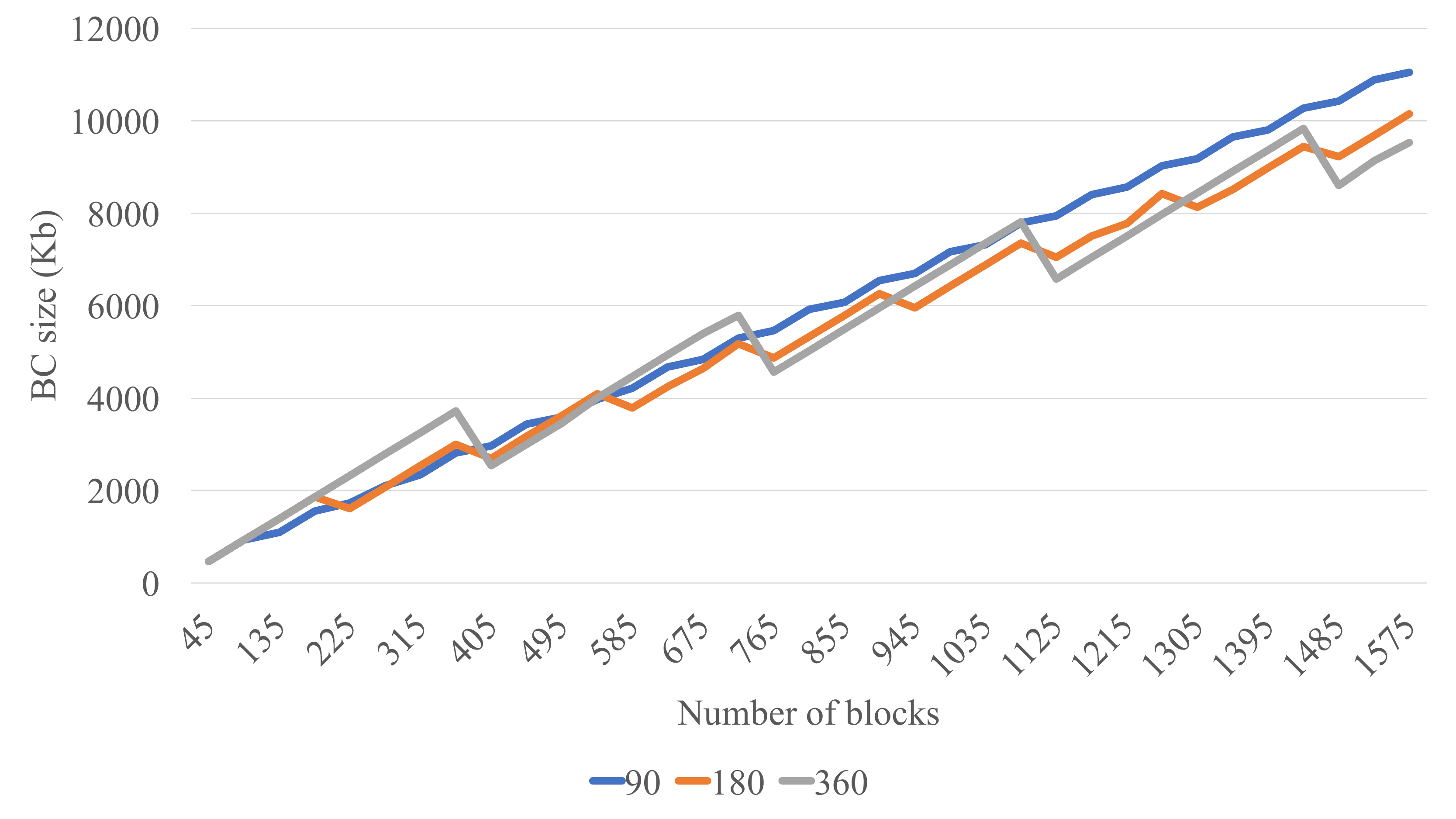}}
	\caption{The impact of the CP on the BC size.}
	\label{fig:evaluation-bcsizevscp}
\end{figure}

 
The cleaning  process (see Section \ref{sec:cleaning}) and summarization (see Section \ref{sec:network-initiated}) tasks performed at the end of each CP directly affect the BC size. Thus, we first  measure the cumulative size of the BC while varying the  CP value. In this setup, the participating nodes are grouped in three groups, each with 300 nodes. The nodes in each group generate transactions with a particular MOM, i.e., permanent, temporary, and summarizable. The nodes that generate temporary transactions are further divided in three groups, each generating transactions with different TTL that  highlights the impact of TTL in BC size. The TTL values equal to   26,  52, and 104 weeks.  This configuration is referred to as the default configuration in the rest of this section.   Figure \ref{fig:evaluation-bcsizevscp} plots the BC size for 3 different CP values, which are 90, 180, and 360 weeks, as a function of the total number of blocks generated and stored in the BC. As can be inferred, with larger CP  more  transactions are collected   which  increases the size of the BC before reaching the CP. Recall from Section \ref{sec:network-initiated} that at the end of each CP, the SMA stores new summary transactions that can no longer be optimized to free up BC memory. Thus, with smaller CPs the frequency which the summary transactions are generated increases. Consequently, the amount of freed up space in the smaller CPs is smaller than larger CPs. \par  
\begin{figure}[h]
	\centerline{\includegraphics[width=9cm,height=6cm,keepaspectratio]{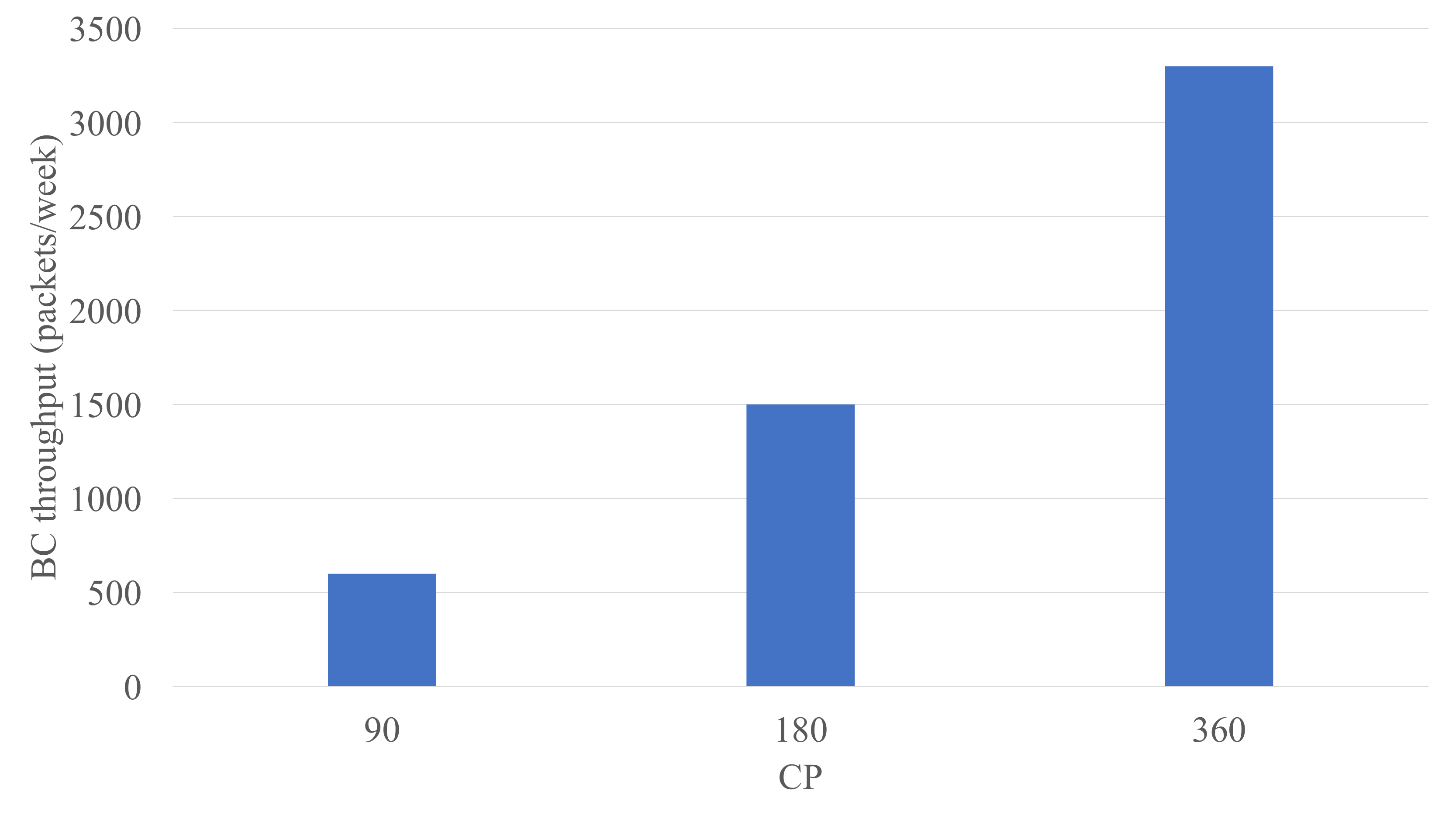}}
	\caption{The impact of CP on the BC throughput.}
	\label{fig:evaluation-throughput}
\end{figure}

\begin{figure}[h]
	\centerline{\includegraphics[width=9cm,height=6cm,keepaspectratio]{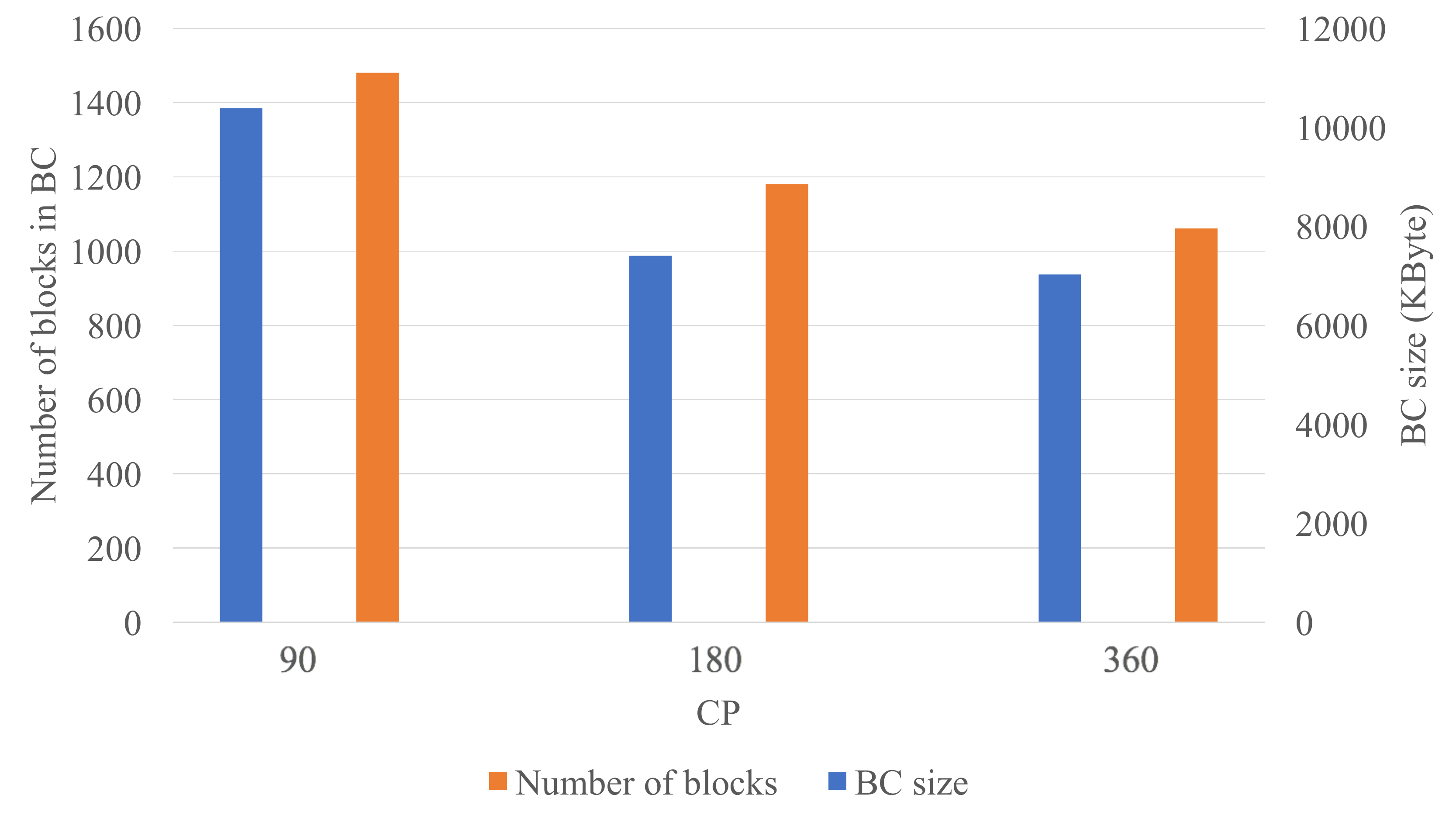}}
	\caption{The impact of CP on the size and length of BC for storing 10k  transactions.}
	\label{fig:evaluation-100transactions}
\end{figure}

At the end of the CP, the summary transactions must be mined into the BC. Consequently,   the number of  transactions generated by the users that can be mined during the CP, referred to as throughput in the rest of this section, is reduced.  Next, we study the impact of varying the CP value on the BC throughput.  We used the default configuration  and the results are shown in  Figure \ref{fig:evaluation-throughput}. As expected, by increasing the CP the BC throughput is also increases.  The CP impacts the number of summary transactions and the BC throughput. These   affect the BC size and number of blocks in BC, as a number of blocks have to be mined into the BC to store the summary transactions. The number of blocks in the BC further impacts the packet and processing overhead on the network for mining and broadcasting the new blocks.  In this part of our evaluations, we measure  the BC size and the number of blocks for storing   a particular number of transactions, 10,000 in our implementations, in the BC. The BC size metric shows the impact  of  summary transactions  on the BC memory footprint.   We use the default configuration and measure the two metrics when 10,000  transactions generated by the users are mined into the BC.   Figure \ref{fig:evaluation-100transactions} presents the  results. It can be seen that for storing the same number of transactions, a larger CP results in a lower memory footprint. Additionally, fewer blocks are required to store  10,000 transactions. Arguably,  the  mining overhead as well as  the bandwidth consumption are decreased and the scalability is increased as fewer blocks need to be broadcast in the network. \par

\begin{figure}[h]
	\centerline{\includegraphics[width=9cm,height=6cm,keepaspectratio]{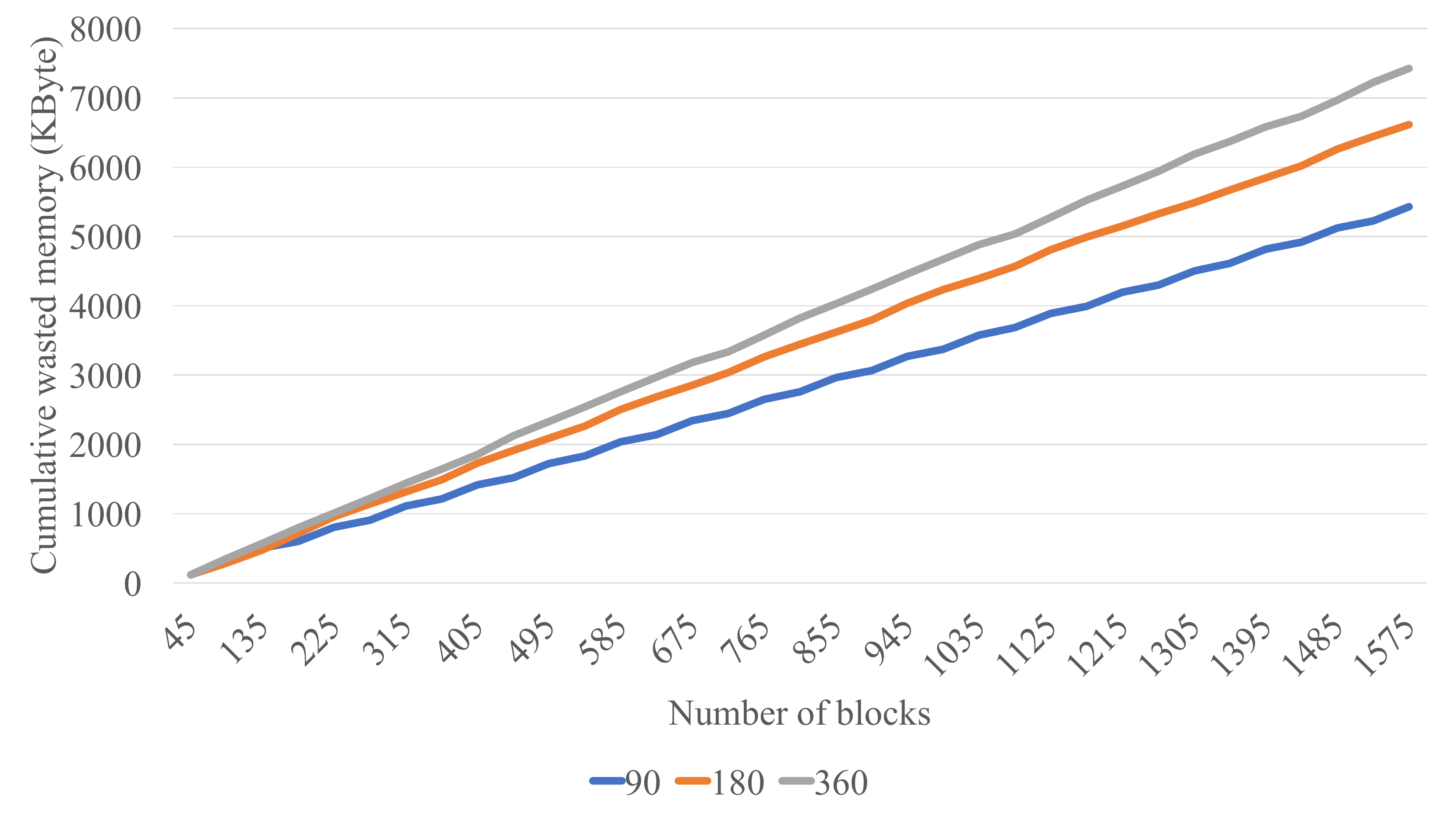}}
	\caption{Evaluation on cumulative wasted memory.}
	\label{fig:evaluation-wastememory}
\end{figure}

Our results have so far shown that larger CPs improve  the BC throughput and memory requirement  compared to the smaller CPs.  However, using large CP incurs  storage overhead at the miners for storing expired transactions including temporary transactions whose  TTL has passed and summarizable transactions,  which is referred to as wasted memory in the rest of the paper, for a longer period.  The main aim of generating the summarizable transactions is to allow them to be removed and thus optimize the BC storage. Thus, they can be considered as wasted memory while they are not yet summarized. To measure the impact of the CP on the wasted memory, we consider the default configuration.   Figure \ref{fig:evaluation-wastememory} plots the cumulative  amount of wasted memory in each miner as a function of the number of blocks for different CPs. With a large CP, the time between when the cleaning process is executed is longer and thus a greater number of expired transactions accumulate. Consequently the wasted memory is far greater than with a smaller CP.    \par 

The wasted memory increases the monetary cost of the miners for storing wasted space. To evaluate the impact of CP on the cost, we measure  the cumulative cost each miner incurs in storing  wasted memory. We based our measurements on the estimated cost in section \ref{sec:network-initiated}, i.e., 0.00000065\$ for five years. The wasted cost can be measured by multiplying the wasted memory given in Figure \ref{fig:evaluation-wastememory} by 0.000000065\$ that is the cost of storing one page of data, i.e., 1KByte,  for six months (which equals with 45 blocks). The latter is the base when we measured the wasted memory. \par 
In conclusion, for choosing a CP the trade-off between throughput memory footprint on the one side and wasted memory and monetary cost on the other needs to be considered.

\section{Conclusion} \label{sec:conclusion}
The immutable nature of the BC  makes it impossible to modify or remove a previously stored block and thus increases BC security. However, it leads to storage, privacy, and cost challenges for BC users particularly in large scale networks like Internet of Things (IoT). This paper, proposed a Memory Optimized and Flexible BC (MOF-BC) that empowers users and Service Providers (SPs) to remove a previously stored transaction or  reduce its size by summarizing transactions or aging the data in transactions. The user/SP may decide to offload the associated overheads for optimizing BC memory to the network using Network-Initiated Memory Optimization (NIMO).  To encourage users/SPs to  employ memory optimization, MOF-BC offers flexible transaction fees and rewards.  The MOF-BC introduces a Generator Verifier (GV) which  addresses  key management for large scale networks while maintains the user/SP privacy. Security analysis  show the robustness of  the MOF-BC  against several attacks.  Implementation results show that MOF-BC achieves lower memory consumption while incurring a small processing overhead. \par 
\section{References}
\bibliography{mybibfile}

\end{document}